\documentclass[sigconf]{acmart}

\usepackage{booktabs} % For formal tables
\usepackage{tabularx} % For column/text width tables
\usepackage[english]{babel}
\usepackage{mathrsfs}
\usepackage{multirow}
\usepackage{multicol}\usepackage{array}
\usepackage{arydshln}

\usepackage{amsmath}
\usepackage{amssymb}
\usepackage{booktabs}
\usepackage[inline]{enumitem}
\usepackage{siunitx}
\usepackage{stackengine}
\usepackage{extarrows}

\usepackage{caption}
\usepackage{subcaption}
\usepackage{footnote}
\makesavenoteenv{tabular}
\makesavenoteenv{table}

\usepackage{xspace} % insert space when needed.

\newcommand{\specialcell}[2][c]{\begin{tabular}[#1]{@{}c@{}}#2\end{tabular}}

\newcolumntype{L}[1]{>{\raggedright\arraybackslash}p{#1}}
\newcolumntype{C}[1]{>{\centering\arraybackslash}p{#1}}
\newcolumntype{R}[1]{>{\raggedleft\arraybackslash}p{#1}}

\copyrightyear{2020}
\acmYear{2020}
\setcopyright{acmcopyright}
\acmConference[SIGIR '20]{Proceedings of the 43rd International ACM SIGIR Conference on Research and Development in Information Retrieval}{July 25--30, 2020}{Virtual Event, China}
\acmBooktitle{Proceedings of the 43rd International ACM SIGIR Conference on Research and Development in Information Retrieval (SIGIR '20), July 25--30, 2020, Virtual Event, China}
\author{Hamed Zamani, Bhaskar Mitra, Everest Chen, Gord Lueck, Fernando Diaz, Paul N. Bennett,\\Nick Craswell, and Susan T. Dumais} % order will change!
\affiliation{%
  \institution{Microsoft}
}
\email{{hazamani, bmitra, yuxche, gordonl, fdiaz, pauben, nickcr, sdumais}@microsoft.com}

\newcommand{\metis}{clarification pane\xspace}
\newcommand{\model}{RLC\xspace}

\begin{document}

\title{Analyzing and Learning from User Interactions for\\Search Clarification}

\begin{abstract}
Asking clarifying questions in response to search queries has been recognized as a useful technique for revealing the underlying intent of the query. Clarification has applications in retrieval systems with different interfaces, from the traditional web search interfaces to the limited bandwidth interfaces as in speech-only and small screen devices. Generation and evaluation of clarifying questions have been recently studied in the literature. However, user interaction with clarifying questions is relatively unexplored. In this paper, we conduct a comprehensive study by analyzing large-scale user interactions with clarifying questions in a major web search engine. In more detail, we analyze the user engagements received by clarifying questions based on different properties of search queries, clarifying questions, and their candidate answers. We further study click bias in the data, and show that even though reading clarifying questions and candidate answers does not take significant efforts, there still exist some position and presentation biases in the data. We also propose a model for learning representation for clarifying questions based on the user interaction data as implicit feedback. The model is used for re-ranking a number of automatically generated clarifying questions for a given query. Evaluation on both click data and human labeled data demonstrates the high quality of the proposed method.

\end{abstract}
% \keywords{TBW.}
\maketitle

\vspace{-0.2cm}
\section{Introduction}
\label{sec:intro}
Search queries are oftentimes ambiguous or faceted. The information retrieval (IR) community has made significant efforts to effectively address the user information needs for such queries. A general approach for obtaining more accurate query understanding is to utilize contextual information, such as short- and long-term interaction history~\cite{Bennett:2012,Kong:2015,Matthijs:2011,Ustinovskiy:2013} and situational context~\cite{Kharitonov:2012,Zamani:2017:WWW}. However, contextual features do not always help the system reveal the user information needs~\cite{Sanderson:2008}. An alternative solution is diversifying the result list and covering different query intents in the top ranked documents~\cite{Santos:2015}. Although result list diversification has been successfully deployed in modern search engines, it still can be a frustrating experience for the users who have to assess the relevance of multiple documents for satisfying their information needs~\cite{Aliannejadi:2019}. On the other hand, in the search scenarios with limited bandwidth user interfaces, presenting a result list containing multiple documents becomes difficult or even impossible~\cite{Aliannejadi:2019,Zamani:2020:WWW}. These scenarios include conversational search systems with speech-only or small screen interfaces. To address these shortcomings, (conversational) search engines can \emph{clarify} the user information needs by asking a question, when there is an uncertainty in the query intent. %\pnb{Uncertainty in search quality or query intent? It seems mainly the latter to me}

Although generating plausible clarifying questions for open-domain search queries has been one of a long-standing desires of the IR community~\cite{Belkin:1995}, it has not been possible until recently. \citet{Zamani:2020:WWW} has recently proposed a neural sequence-to-sequence model that learns to generate clarifying questions in response to open-domain search queries using weak supervision. They showed that clarifying questions can be of significance even for web search engines with the traditional ten blue link interface. 

\begin{figure*}[t]
\vspace{-0.6cm}
\begin{subfigure}{.33\textwidth}
  \centering
  \includegraphics[width=\linewidth,trim={5cm 34cm 42cm 6cm},clip]{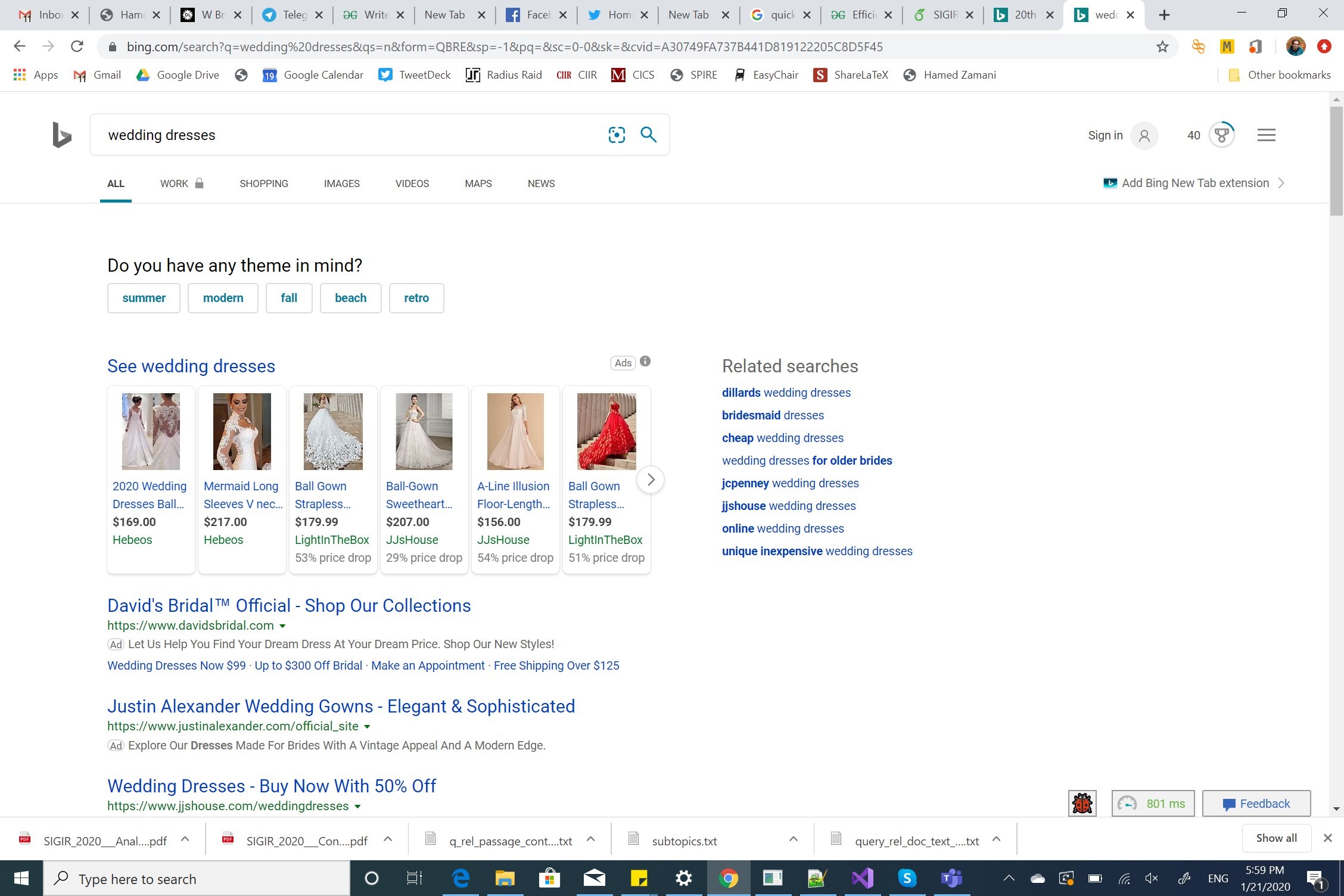}
%   \caption{\# candidate answers = 2}
%   \label{fig:sfig1}
\end{subfigure}%
\begin{subfigure}{.33\textwidth}
  \centering
  \includegraphics[width=\linewidth,trim={5cm 34cm 42cm 6cm},clip]{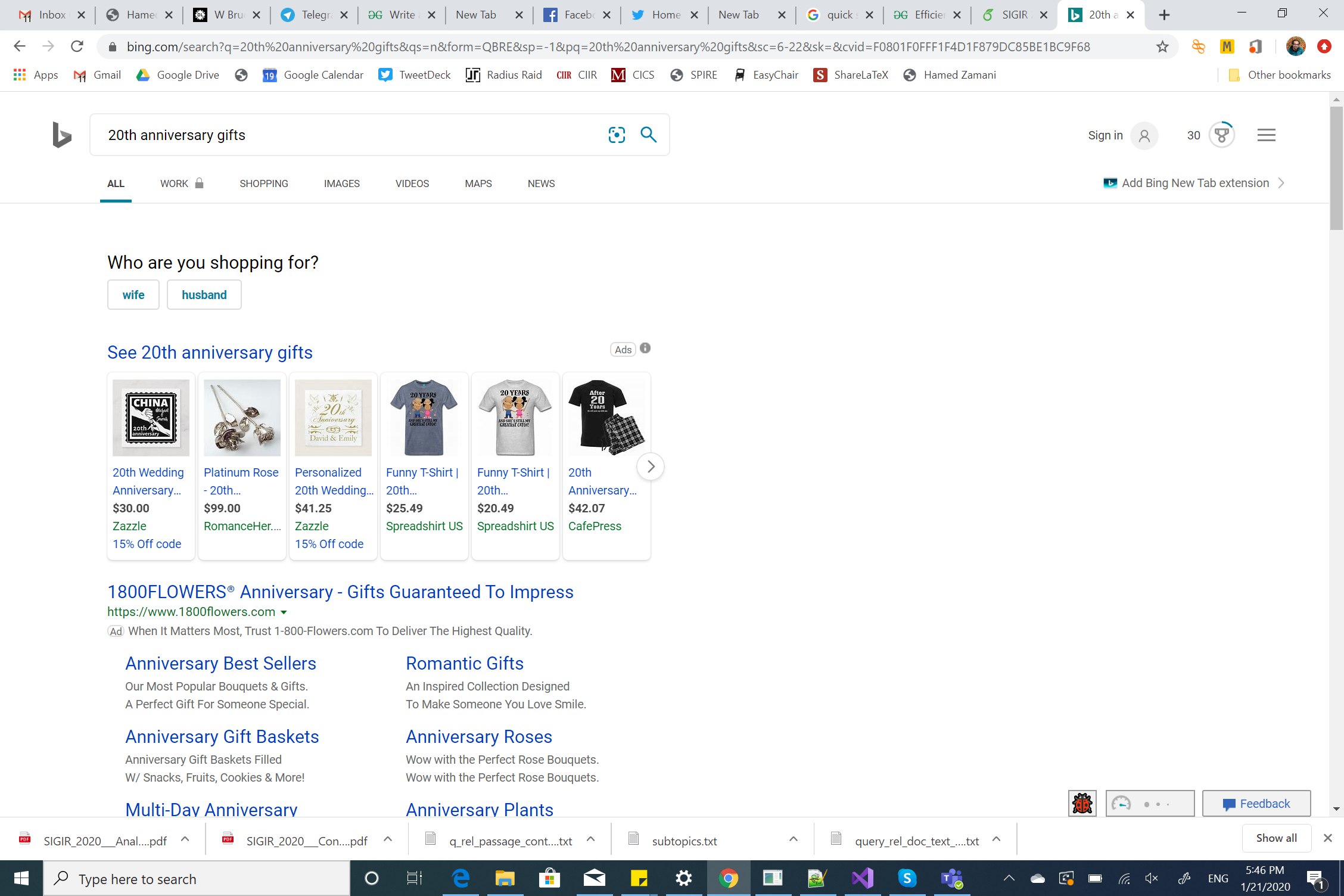}
%   \caption{\# candidate answers = 3}
%   \label{fig:sfig1}
\end{subfigure}%
\begin{subfigure}{.33\textwidth}
  \centering
  \includegraphics[width=\linewidth,trim={5cm 34cm 42cm 6cm},clip]{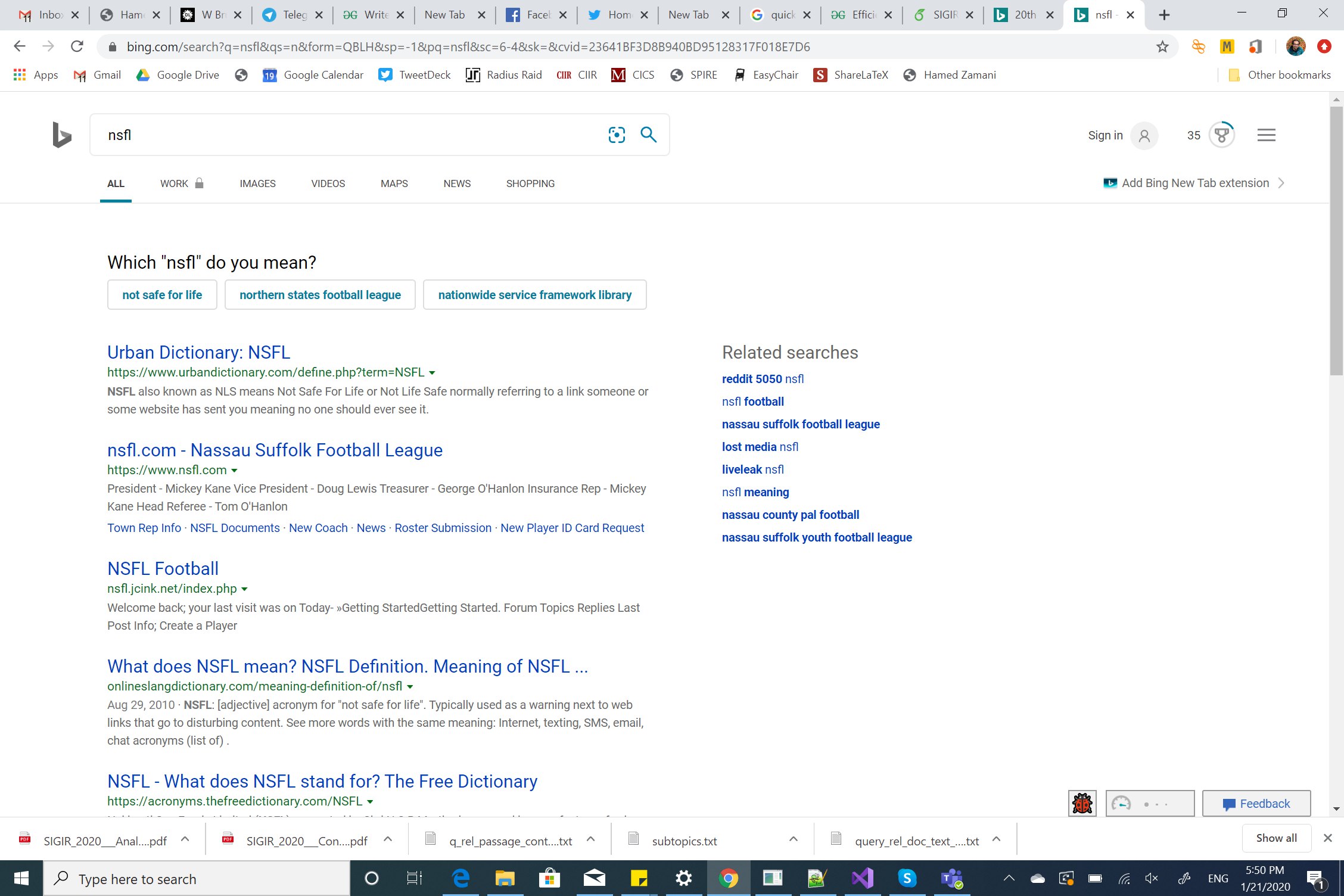}
%   \caption{\# candidate answers = 4}
%   \label{fig:sfig2}
\end{subfigure}%
\caption{Few examples of clarification in web search.}
\label{fig:bing}
% \vspace{-0.4cm}
\end{figure*}

Despite the significant progress in exploring clarification in search~\cite{Aliannejadi:2019,Zamani:2020:WWW} and related areas~\cite{Mostafazadeh:2016,Rao:2018,Trienes:2019}, the way users interact with such conversational features of search engines is relatively unknown. Analyzing user interactions with clarifying questions would lead to a better understanding of search clarification, and help researchers realize which queries require clarification and which clarifying questions are preferred by users. Based on this motivation, we conduct a large-scale study of user interactions with clarifying questions for millions of unique queries. This study is based on a relatively new feature, called \emph{\metis}, in the Bing search engine that asks a clarifying question in response to some queries. %\footnote{This feature is not available for some international markets.} 
The interface is shown in \figurename~\ref{fig:bing}. We analyze user engagements with clarifying questions based on different attributes of the \metis. We also study user interactions with clarifying questions for different query properties, such as query length and query type (natural language question or not, ambiguous or faceted, tail or head). We further perform a preliminary study on click bias in clarification panes. Our comprehensive analyses lead to a number of suggestions for improving search clarification.

Following our user interaction analyses, we propose a model for learning representations for clarifying questions together with their candidate answers from user interactions as implicit feedback. Our model consists of two major components: Intents Coverage Encoder and Answer Consistency Encoder. The former encodes the intent coverage of the \metis, while the latter encodes the plausibility of the \metis, i.e., the coherency of the candidate answers and their consistency with the clarifying questions. Our model is solely designed based on the attention mechanism. We evaluate the model using click data as well as human labeled data. The experiments suggest significant improvements compared to competitive baselines.

In summary, the major contributions of this work include:
% \vspace{-0.2cm}
\begin{itemize}[leftmargin=*]
    \item Conducting the first large-scale analysis of user interactions with clarification panes in search. Our study provides suggestions for the future development of algorithms for search clarification.
    \item Performing preliminary experiments showing different click biases, including both position and presentation biases, in the user interaction data with clarification.
    \item Proposing a novel neural model, specifically designed for representation learning for clarifying questions. Our model outperforms competitive baselines for the task of clarifying question selection/re-ranking.
\end{itemize}

\vspace{-0.3cm}
\section{Related Work}
\label{sec:related}
In this section, we review prior work on asking clarifying questions, query suggestion, and click bias estimation.

\vspace{-0.2cm}
\paragraph{\textbf{Asking Clarifying Question}}
Clarifying questions have been found useful in a number of applications, such as speech recognition~\cite{Stoyanchev:2014} as well as dialog systems and chat-bots~\cite{DeBoni:2003,DeBoni:2005,Quintano:2008}. In community question answering websites, users often use clarifying questions to better understand the question~\cite{Braslavski:2017,Rao:2018,Rao:2019}.
\citet{Kiesel:2018} studied the impact of voice query clarification on user satisfaction. They concluded that users like to be prompted for clarification. \citet{Coden:2015} studied clarifying questions for entity disambiguation mostly in the form of ``did you mean A or B?''. Recently, \citet{Aliannejadi:2019} suggested an offline evaluation methodology for asking clarifying questions in conversational systems by proposing the Qulac dataset. The importance of clarification has been also discussed by \citet{Radlinski:2017}. In the TREC HARD Track~\cite{Allan:2004}, participants could ask clarifying questions by submitting a form in addition to their runs. Most recently, \citet{Zamani:2020:WWW} proposed models for generating clarifying questions for open-domain search queries. In another study, \citet{Zamani:2020:Macaw} developed a platform for conversational information seeking that supports mixed-initiative interactions, including clarification. In addition, \citet{Hashemi:2020:SIGIR} introduced a neural model for representing user interactions with clarifying questions in an open-domain setting. Asking clarifying questions about item attributes has been also explored in the context of conversational recommender systems~\cite{Sun:2018}. For instance, \citet{Christakopoulou:2016} designed a system for preference elicitation in venue recommendation. \citet{Zhang:2018} automatically extracted facet-value pairs from product reviews and considered them as questions and answers. In contrast to prior work on search clarification, this work focuses on understanding user interactions with clarifying questions in a real system based on log analysis.

\vspace{-0.2cm}
\paragraph{\textbf{Query Suggestion and Auto-Completion}} 
Query suggestion techniques~\cite{Dehghani:2017,Ozertem:2012,Santos:2013} are used to suggest useful next queries to the users. They have been successfully implemented in search engines. Query suggestion, although related, is fundamentally different from search clarification. The reason is that candidate answers should clarify the intent behind the current search query. While, in query suggestion, the next search query might be a follow up query that is often searched after the query. The clarification examples presented in \figurename~\ref{fig:bing} clearly show the differences. The provided candidate answers are not the expected query suggestions. 

Query auto-completion, on the other hand, makes suggestion to complete the current search query~\cite{Cai:2016,Mitra:2014,Shokouhi:2013}. In contrast to query auto-completion, search clarification asks a clarifying question and provides coherent candidate answers which are also consistent with the clarifying question. For more details on the differences between search clarification and query suggestion or auto-completion, we refer the reader to \cite{Zamani:2020:WWW}.

\vspace{-0.2cm}
\paragraph{\textbf{Click Bias}}
Click bias in user interactions with search engines has been extensively explored in the literature. It has been shown that users intend to click more on the documents with higher rank positions. There exist different biases, such as position bias~\cite{Joachims:2005}, presentation bias~\cite{Yue:2010}, and trust bias~\cite{Agarwal:2019}. To address this issue, several user models for simulating user behavior have been proposed, such as the Examination model~\cite{Richardson:2007} and the Cascade model~\cite{Craswell:2008}. In our clarification interface, the candidate answers are presented horizontally to the users. The answer length is also short, thus multiple answers can be seen at a glance. These unique properties make the click bias in clarification different from document ranking. It is even different from image search, in which the results are shown in a two dimensional grid interface~\cite{Ohare:2016,Xie:2018}. 

%There exist several user models that are out of the scope of this paper.

%  \hamed{complete this}

% \paragraph{\textbf{Neural Models for IR}} \hamed{do we really need this?}

\vspace{-0.2cm}
\section{Analyzing User Interactions with Clarification}
\label{sec:analysis}

In this section, we study user interactions with clarifying questions in Bing, a major commercial web search engine. We believe these analyses would lead to better understanding of user interactions and expectations from search clarification, which smooths the path towards further development and improvement of algorithms for generating and selecting clarifying questions. 

In the following subsections, we first introduce the data we collected from the search logs for our analyses. We further introduce the research questions we study in the analyses and later address these questions one by one.

% \begin{figure}[t]
%     \centering
%     \includegraphics[width=\linewidth]{img/bing2.png}
%     % \vspace{-0.5cm}
%     \caption{An example of clarification in web search. \hamed{find a better example}}%\protect\footnotemark}
%     \label{fig:bing}
%     % \vspace{-0.5cm}
% \end{figure}

% \begin{table}[t]
%     \centering
%     \caption{Statistics of the data collected from the user interactions with the \metis.}
%     \begin{tabular}{ll}\toprule
%         Total impressions & 74,617,653 \\
%         Total clicks & 810,723 \\
%         \# unique query-clarification pairs & 12,344,924 \\
%         \# unique query-clarification pairs with 10+ impressions & 956,151 \\
%         \# unique query-clarification pairs with positive CTR & 450,694 \\
%         \# unique queries & 5,553,850 \\
%         \# unique queries with multiple clarification panes & 2,302,532 \\
%         Average number of clarification panes per query & $3.94 \pm 7.28$ \\
%         Average CTR & 1.00\% \\
%         Average number of candidate answers & $2.99 \pm 1.14$ \\
%         \bottomrule
%     \end{tabular}
%     \label{tab:metis_data}
% \end{table}

\begin{table}[t]
    \centering
    \caption{Statistics of the data collected from the user interactions with the \metis.}
    \vspace{-0.4cm}
    \begin{tabular}{ll}\toprule
        Total impressions & 74,617,653 \\
        \# unique query-clarification pairs & 12,344,924 \\
        \# unique queries & 5,553,850 \\
        \# unique queries with multiple clarification panes & 2,302,532 \\
        Average number of candidate answers & $2.99 \pm 1.14$ \\
        \bottomrule
    \end{tabular}
    \label{tab:metis_data}
    \vspace{-0.4cm}
\end{table}

\vspace{-0.2cm}
\subsection{Data Collection}
\label{sec:analysis:data}
The search engine asks clarifying questions from users in response to some ambiguous or faceted queries. The user interface for this feature, which is called the \emph{\metis}, is shown in \figurename~\ref{fig:bing}. The \metis is rendered right below the search bar and on top of the result list. Its location in the result page never changes. The \metis consists of a clarifying question and up to five clickable candidate answers. 
% Due to the production standards in commercial search engines, in case the system cannot generate a specific question with a high confidence score, it produces a generic question. 
% The number of candidate answers presented to the users varies between two and five inclusively, depending on the query. 
Note that the \metis is not triggered for navigational queries.
To conduct the analyses, we obtained the clickthrough data for the \metis in Bing. For some queries, the data contains multiple clarification panes shown to different set of users. The difference between these clarification panes relies on the clarifying question, the candidate answer set, or even the order of candidate answers. For more information on generating clarification panes, we refer the reader to \cite{Zamani:2020:WWW}.

% We monitored and logged the clickthrough data for the \metis in Bing. To conduct a deep analysis of user interactions with the \metis, for some queries we generated multiple different clarification panes. The difference between the clarification panes might be in the question, in the candidate answer set, or even in their ordering. We performed multiple different randomization to cover various analyses. Note that all the generated clarification panes satisfy the production standards and are still with high quality. 

The collected data consists of over 74.6 million \metis impressions (i.e., the number of times the \metis was shown to users). The data consists of over 5.5 million unique queries. The average number of candidate answers per \metis is equal to 2.99. The statistics of the data is reported in \tablename~\ref{tab:metis_data}. Note that we only focus on the query-clarification pairs with at least 10 impressions.

% \begin{figure}
% \begin{subfigure}{.25\textwidth}
%   \centering
%   \includegraphics[width=\textwidth]{img/log_log_total_engagement.png}
%   \caption{}
%   \label{fig:ctr_dist:a}
% \end{subfigure}%
% \begin{subfigure}{.25\textwidth}
%   \centering
%   \includegraphics[width=\linewidth]{img/ctr-dist-log-scale.pdf}
%   \caption{}
%   \label{fig:ctr_dist:b}
% \end{subfigure}%
% %
% \caption{(a) A log-log plot of total engagement versus rank in the clarification data, and (b) the engagement rate distribution in a logarithmic space.}
% \label{fig:ctr_dist}
% \end{figure}

% \begin{figure}[t]
%     \centering
%     \includegraphics[width=\linewidth]{img/ctr-dist-log-scale.pdf}
%     \caption{Total CTR distribution in log scale. Minimum impression: 10.}
%     \label{fig:ctr_dist}
% \end{figure}

% \begin{figure}[t]
%     \centering
%     \includegraphics[width=\linewidth]{img/log_log_total_engagement.pdf}
%     \caption{Total CTR distribution in log scale. Minimum impression: 10.}
%     \label{fig:ctr_dist}
% \end{figure}

\vspace{-0.2cm}
\subsection{Research Questions}
\label{sec:analysis:rqs}
In the rest of Section~\ref{sec:analysis}, we study the following research questions by analyzing the user interaction data described in Section~\ref{sec:analysis:data}.

\begin{enumerate}
    % \item[\textbf{RQ1}] How often do users interact with clarifying questions in web search? (Section~\ref{sec:analysis:rq1}).
    \item[\textbf{RQ1}] Which clarifying questions would lead to higher engagements? (Section~\ref{sec:analysis:rq2}).
    \item[\textbf{RQ2}] For which search queries do users prefer to use clarification? (Section~\ref{sec:analysis:rq3})
    \item[\textbf{RQ3}] How is the impact of clarification on search experience? (Section~\ref{sec:analysis:rq4}) %\hamed{not answered yet! waiting for the data!}
\end{enumerate}

% \subsection{Analyzing the Overall Clarification Engagement}
% \label{sec:analysis:rq1}
% To study the first research question described above (RQ1), we study the overall user engagements with the \metis in our data. We sorted the query-clarification pairs with respect to their overall click count (as a major measure for user engagement) in descending order, and present a log-log plot of total engagement versus rank in \figurename~\ref{fig:ctr_dist:a}. The plot highlights that these quantities follow the power law, similar to the Zipfian distribution. In other words, there exist few clarification panes with high click count and there is a long tail in the distribution. The next plot (\figurename~\ref{fig:ctr_dist:b}), on the other hand, focuses on engagement rate (i.e., clickthrough rate) and shows its distribution in the clarification data. For better visualization, the distribution is plotted in the logarithmic space. According to the plot, most clarifications receive less than 5\% engagement. Note that the \metis is presented on top of the search engine result page (SERP) and most users skip the \metis and directly go to the result list. The average engagement of $1\%$ is still relatively successful. The distribution for those with less than $40\%$ engagement is close to the exponential distribution (linear in the logarithmic space) and for those with higher engagement (especially those with over $70\%$ engagements), it gets closer to the uniform distribution.

\begin{figure}[t]
    \centering
    \vspace{-0.6cm}
    \includegraphics[width=.65\linewidth]{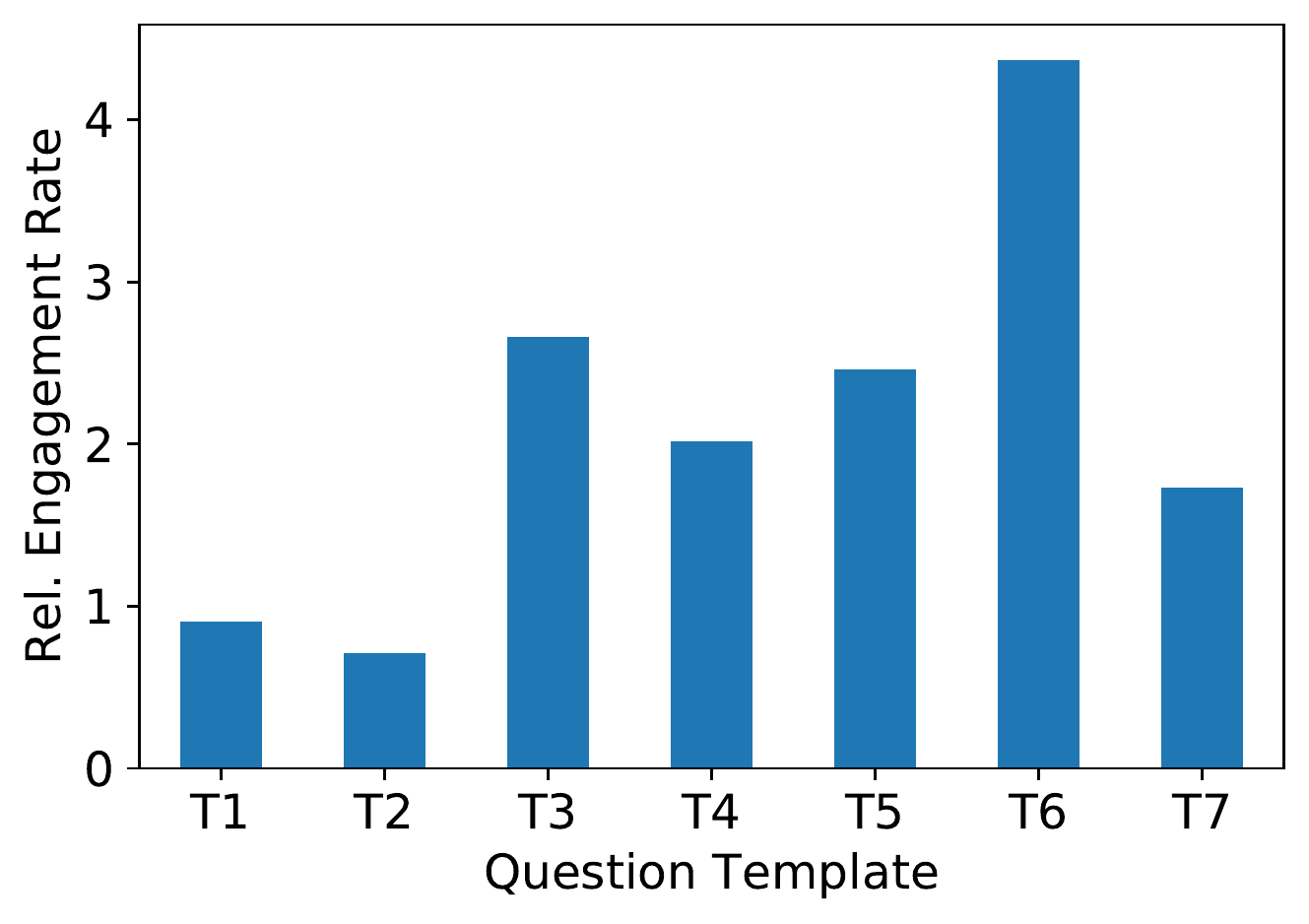}
    \begin{minipage}{1.0\textwidth}
    T1: What (would you like | do you want) to know about \_\_\_\_\_? \\
    T2: (Which | What) \_\_\_\_\_ do you mean? \\
    T3: (Which | What) \_\_\_\_\_ are you looking for? \\
    T4: What (would you like | do you want) to do with \_\_\_\_\_? \\
    T5: Who are you shopping for? \\
    T6: What are you trying to do? \\
    T7: Do you have \_\_\_\_\_ in mind?
    \end{minipage}
    \vspace{-0.4cm}
    \caption{Relative engagement rate (compared to the average engagement rate) per question template for the most frequent templates in the data.}
    \label{fig:ctr_per_question}
    \vspace{-0.6cm}
\end{figure}

\vspace{-0.2cm}
\subsection{Characterizing Clarifications with High Engagement Rate}
\label{sec:analysis:rq2}
In this subsection, we address RQ1 defined in Section~\ref{sec:analysis:rqs}. To this end, we study the obtained engagement rate (i.e., click rate) by the \metis based on different clarification properties, including (1) the clarifying question template, (2) the number of candidate answers, and (3) the conditional click distribution across candidate answers.

\vspace{-0.2cm}
\subsubsection{Analyzing Clarifying Question Templates}
% \pnb{Next sentence seems like an overly strong claim -- ``all'' can be addressed with templates}
As recently discovered by \citet{Zamani:2020:WWW}, most clarification types can be addressed using a set of pre-defined question templates. We identified all question templates used in the data and focused on the most frequent templates. The average engagement rate obtained by each template relative to the overall average engagement rate is presented in \figurename~\ref{fig:ctr_per_question}. The templates are sorted with respect to their reverse frequency in the data. According to the figure, general question templates than can be potentially used for almost all queries, such as ``what would you like to know about QUERY?'' have the higher frequency in the data, while their engagement is relatively low. On the other hand, more specific question templates,\footnote{By more specific, we mean the questions that cannot be asked for all queries, as opposed to general templates, like T1, that can be asked in response to any query.} such as ``what are you trying to do?'', ``who are you shopping for?'', and ``which \_\_\_\_\_ are you looking for?'' lead to much higher engagement rates. 
% \pnb{This is really interesting but where do we describe how we judge general vs.\ specific. These are not axes of the WWW paper taxonomy. } 
The relative difference between the engagement rates received by the templates can be as large as $500\%$ (T2 vs. T6). 

% \hamed{analysis per clarification type in our WWW taxonomy?}

% \begin{figure}[t]
%     \centering
%     \includegraphics[width=\linewidth]{img/avg-ctr-per-num-options.pdf}
%     \caption{Average CTR per number of candidate answers. Minimum impression: 10.}
%     \label{fig:ctr_per_num_options}
% \end{figure}

\begin{table}[t]
    \centering
    \vspace{-0.6cm}
    \caption{Relative engagement rate (w.r.t. average engagement) for clarification panes per number of answers.}
    \vspace{-0.4cm}
    \begin{tabular}{lcccc}\toprule
        \textbf{\# Candidate Answers} & \textbf{2} & \textbf{3} & \textbf{4} & \textbf{5} \\\midrule
        \textbf{Relative Engagement Rate} & 0.95 & 1.05 & 1.03 & 1.03 \\\bottomrule
    \end{tabular}
    \label{tab:ctr_per_num_options}
    \vspace{-0.4cm}
\end{table}

\vspace{-0.2cm}
\subsubsection{Analyzing the Number of Candidate Answers}
As mentioned earlier in Section~\ref{sec:analysis:data}, the number of candidate answers varies between two and five. \tablename~\ref{tab:ctr_per_num_options} shows the relative engagement rate per number of candidate answers in the \metis. According to the results, the clarification panes with only two candidate answers receive a slightly lower engagement rate. The reason could be that the clarification panes with two candidate answers do not always cover all aspects of the submitted query. The clarifying questions with more than two candidate answers generally receive similar engagement rates with each other. Generally speaking, the number of candidate answers is not a strong indicator of user engagement rate.

\begin{figure}[t]
    \centering
    % \vspace{-0.4cm}
    \includegraphics[width=.65\linewidth]{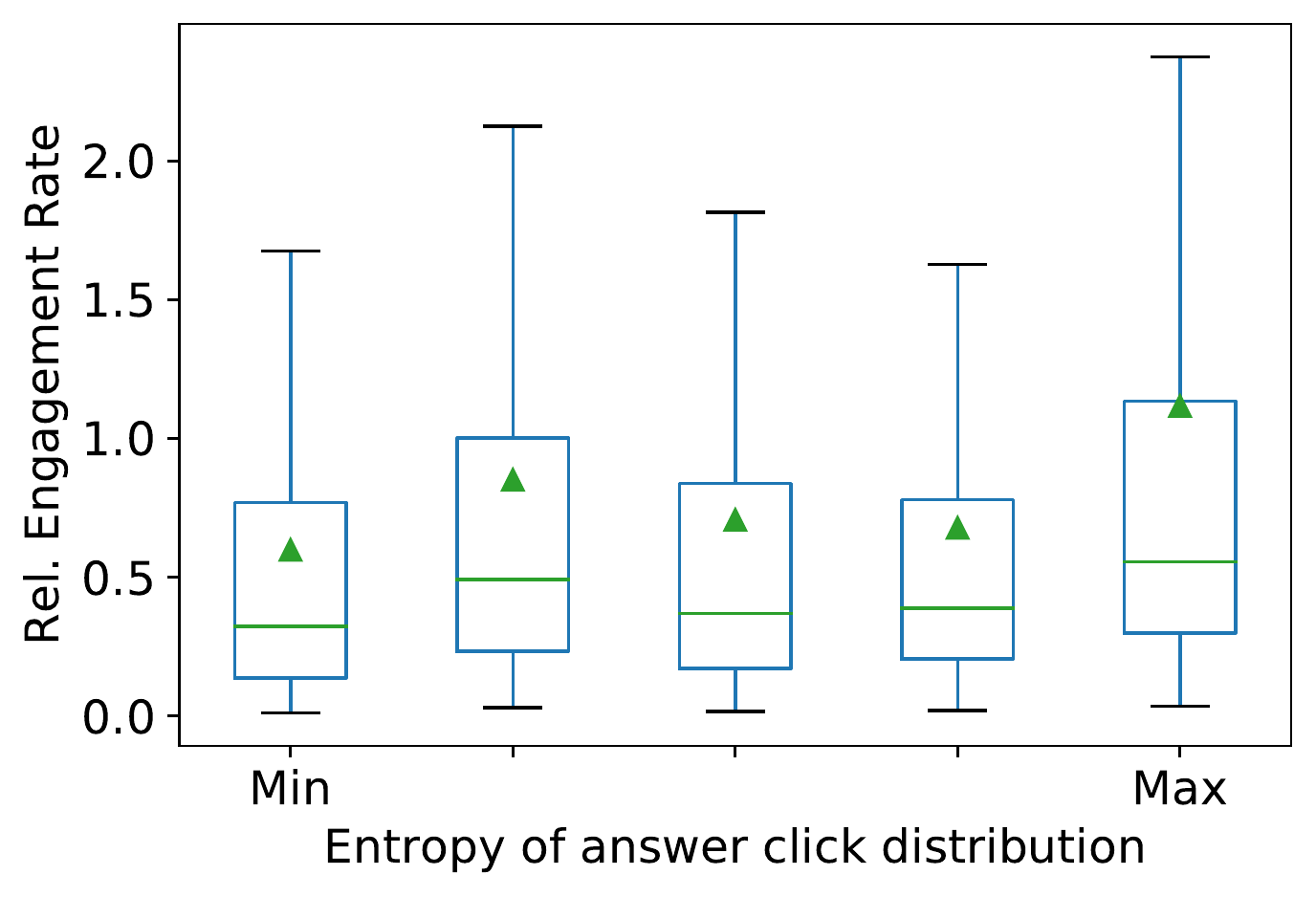}
    \vspace{-0.4cm}
    \caption{A box plot for the relative engagement rate (compared to the average engagement rate) with respect to the entropy in the conditional answer click distribution. This plot is only computed for clarifications with five options.}
    \label{fig:entropy_ctr}
    \vspace{-0.4cm}
\end{figure}

\vspace{-0.2cm}
\subsubsection{Analyzing Answer Click Distribution}
\figurename~\ref{fig:entropy_ctr} plots the relative engagement rate received by the \metis with respect to the entropy of conditional click distribution on the candidate answers. In case of no observed click for a \metis, we assigned equal conditional click probability to all candidate answers. The box plot is computed for five equal-width bins between the minimum and maximum entropy. In this experiments, we only focus on the clarification panes with exactly five candidate answers. The goal of this analysis is to discuss whether higher click entropy (i.e., closer to the uniform click distributions on the candidate answers) would lead to higher engagement rate. According to the plot, the clarification panes with the highest click entropy lead to the highest average and median engagement rate. The second bin from the left also achieves a relatively high average and median engagement rate. This plot shows that the data points in the minimum entropy bin achieve the lowest average and median engagement rates, however, the increase in answer click entropy dose not always lead to higher engagement rate. The reason is that some clarification panes with high engagement rates contain a dominant answer. As an example, for the clarifying question ``What version of Windows are you looking for?'', we observe over 10 times more clicks on ``Windows 10'' compared to the other versions. Note that this may change over time. Analyzing the temporal aspect of click distribution and engagement rate is left for future work. In summary the majority of engagement comes for one of two reasons: (1) high ambiguity in the query with many resolutions (i.e., the high click entropy case); (2) ambiguity but where there is a dominant ``assumed'' intent by users where they only realize the ambiguity after issuing the query (e.g., the mentioned Windows 10 example).

% \hamed{do we need to summarize the answer to RQ2?}
% \pnb{Is it fair to say that the majority of engagement comes for one of two reasons: (1) high ambiguity in the query with many resolutions (this is the high click entropy case); (2) ambiguity but where there is a dominant "assumed" intent by users where users only realize the ambiguity after issuing the query (the windows 10 example). }

\begin{figure}[t]
    \centering
    \vspace{-0.6cm}
    \includegraphics[width=.65\linewidth]{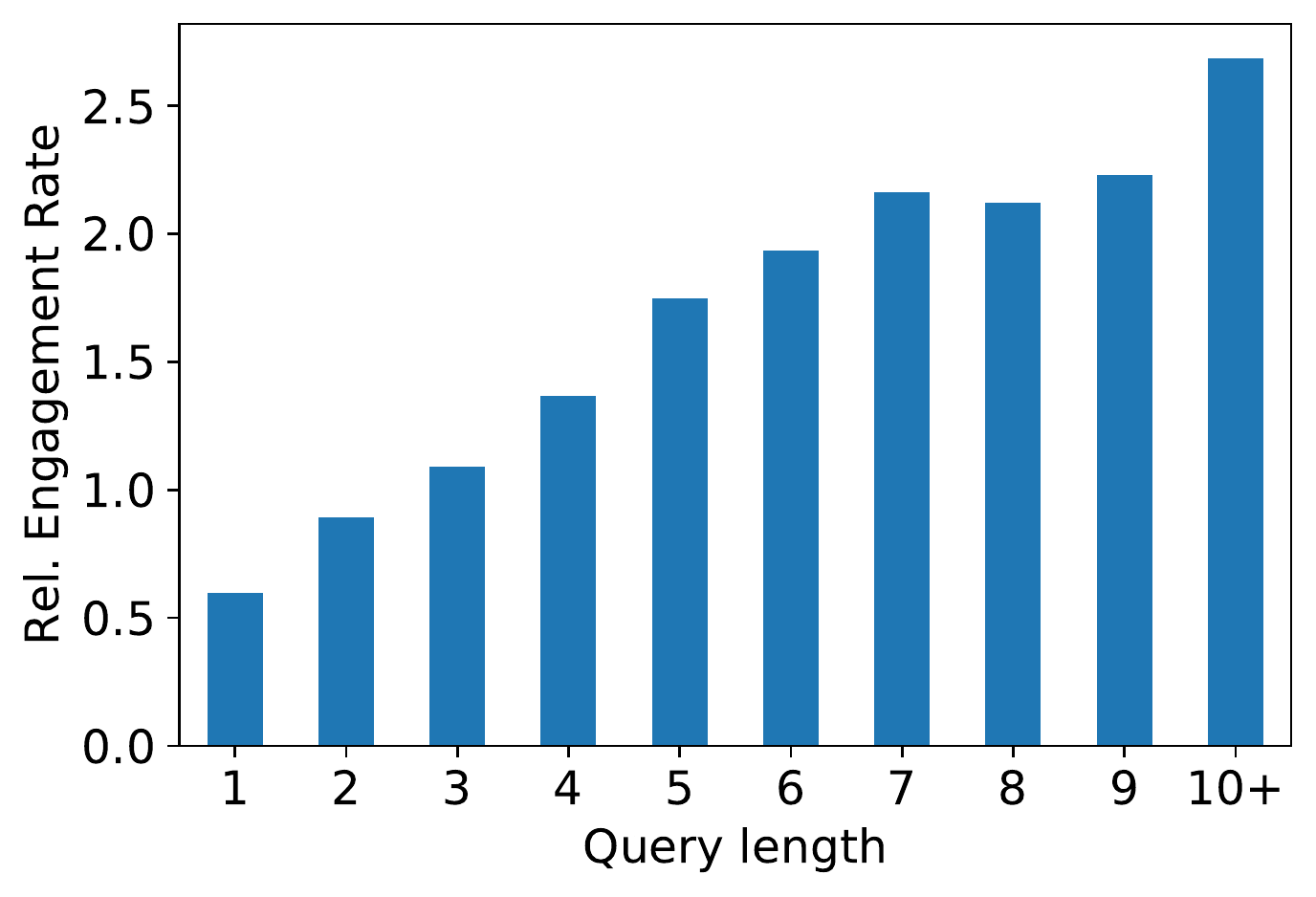}
    \vspace{-0.4cm}
    \caption{Relative engagement rate (compared to the average engagement rate) per query length.}
    \label{fig:ctr_per_query_len}
    \vspace{-0.4cm}
\end{figure}

\vspace{-0.2cm}
\subsection{Characterizing Queries with High Clarification Engagement}
\label{sec:analysis:rq3}
We address the second research question (RQ2: For which web search queries, do users prefer to use clarification?) by analyzing the user engagements with the \metis based on different query properties, such as query length, query type (natural language questions vs. other queries; ambiguous vs. faceted queries; head vs. torso vs. tail queries), and historical clicks observed for the query. 

\vspace{-0.2cm}
\subsubsection{Analyzing Clarification Engagement Based on Query Length}
% \pnb{The ``search quality'' of Bing or clarification quality? Do we need this statement -- it seems like a pretty strong statement about Bing in general rather than ``In the research literature, long queries have often given rise to more challenges in producing quality results''.}
In the research literature, long queries have often given rise to more challenges in producing quality results. One reason is that longer queries are more likely to be less frequent and among tail queries~\cite{Gupta:2015}. We study the engagement rates received by the \metis with respect to the query length. The result is shown in \figurename~\ref{fig:ctr_per_query_len}. Interestingly, as the query length increases, we observe substantial increase in the average engagement rate. Note that the \metis is not shown to the user for navigational queries, thus the data does not contain such queries. 

%There are multiple reasons that can lead to this observation: (1) The search result quality has a negative correlation with the query length. Therefore, users may engage more with clarifying questions when the result list fails to satisfy their information needs. (2) The SERP contains many elements, such as entity card and an extracted answer from the web. These elements are less triggered for longer and tail queries. (3) The users who prefer to write long queries may prefer to engage with such conversational features of the SERP.
%\pnb{In general this section seems to present more questions than it answers and I'm not sure how much I agree with.}

\begin{table}[t]
    \centering
    \vspace{-0.6cm}
    \caption{Relative engagement rate (compared to the average engagement rate) per query type.}
    \vspace{-0.4cm}
    \begin{tabular}{lc}\toprule
        \textbf{Query type} & \textbf{Relative engagement rate} \\\midrule
        Natural language question & 1.58 \\
        Other queries & 0.96 \\\midrule
        Faceted queries & 1.52 \\
        Ambiguous queries & 0.70 \\\midrule
        Tail queries & 1.01 \\
        Torso queries & 1.02 \\
        Head queries & 0.99 \\
        \bottomrule
    \end{tabular}
    \label{tab:ctr_per_query_type}
    \vspace{-0.4cm}
\end{table}

\vspace{-0.2cm}
\subsubsection{Analyzing Clarification Engagement for Natural Language Questions}
According to the first two rows in \tablename~\ref{tab:ctr_per_query_type}, the average engagement rate observed for natural language questions are $64\%$ (relatively) higher than the other queries. Therefore, users who issue natural language questions are more likely to interact with the \metis. This observation demonstrates yet another motivation for using clarifying questions in the information seeking systems with natural language user interactions, such as conversational search systems.
% \pnb{Are clarification questions for NL queries more specific? In which case is it because of the query or the quality of the clarification?}

\begin{figure}[t]
    \centering
    % \vspace{-0.6cm}
    \includegraphics[width=.65\linewidth]{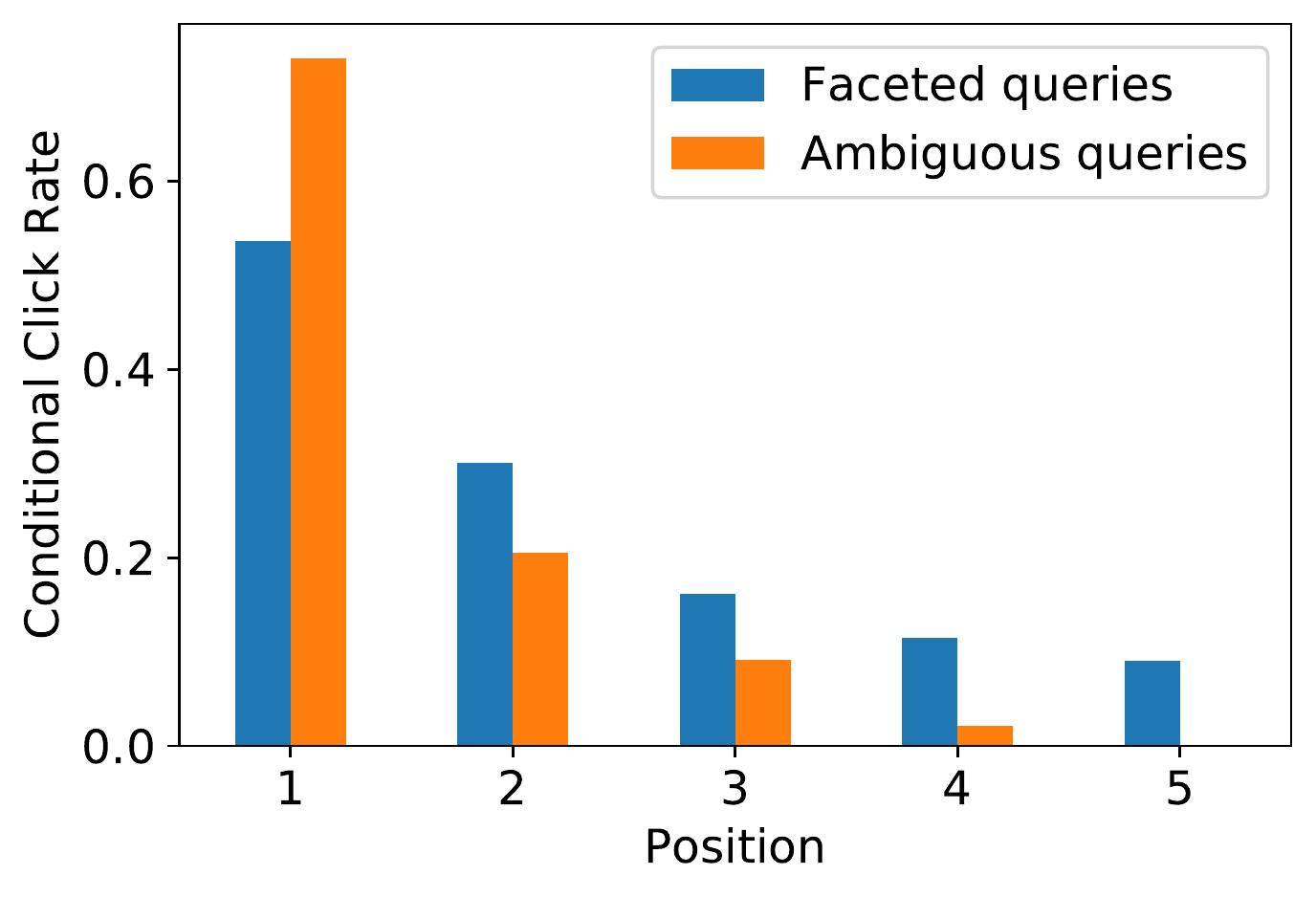}
    \vspace{-0.4cm}
    \caption{Conditional click rate per position for ambiguous vs. faceted queries for clarifications with five answers.} %hamed
    \label{fig:ambiguous_vs_faceted}
    \vspace{-0.4cm}
\end{figure}

\vspace{-0.2cm}
\subsubsection{Analyzing Clarification Engagement for Ambiguous versus Faceted Queries}
Clarifying questions in web search can be useful for revealing the user information needs behind the submitted \emph{ambiguous} or \emph{faceted} queries. In \figurename~\ref{fig:bing}, few clarification examples are shown. The third example in the figure (right) shows the clarification pane for an ambiguous query, while the other two are faceted queries. The middle part of \tablename~\ref{tab:ctr_per_query_type} reports the relative engagement rate received by the \metis for ambiguous and faceted queries. The category of each query was automatically identified based on the clarifying question and the candidate answers generated in the \metis. According to the figure, the \metis for faceted queries are approximately $100\%$ more likely to receive a click compared to the ambiguous queries. We plot the conditional click distribution per position for ambiguous and faceted queries in \figurename~\ref{fig:ambiguous_vs_faceted}. The graph shows that the gap between the first and the second position for ambiguous queries are substantially higher than the gap for faceted queries. This shows that for ambiguous queries, it is more likely that one query intent dominates the user information needs for the query. In fact, this might be one of the reasons that the \metis for ambiguous queries receives less engagement, because it is likely that the SERP often covers the most dominant query intent in the top position, thus users skip the \metis and directly move to the result list.

\vspace{-0.2cm}
\subsubsection{Analyzing Clarification Engagement for Head, Torso, and Tail Queries}
We use the search traffic to identify the query types. The most frequent queries for a third of search traffic was considered as head queries, the second third as torso, and the rest as tail queries. This results in a small number of high frequency head queries and a large number of low frequency tail queries. We further compute the average engagement rate per query types and report the results in the last part of \tablename~\ref{tab:ctr_per_query_type}. According to the results, all query types achieve similar clarification engagement. Note that the data contains the queries that the \metis was triggered for, therefore, there should be too many tail queries that the system does not generate a clarifying question for.

\begin{figure}
\vspace{-0.6cm}
\begin{subfigure}{.25\textwidth}
  \centering
  \includegraphics[width=\textwidth]{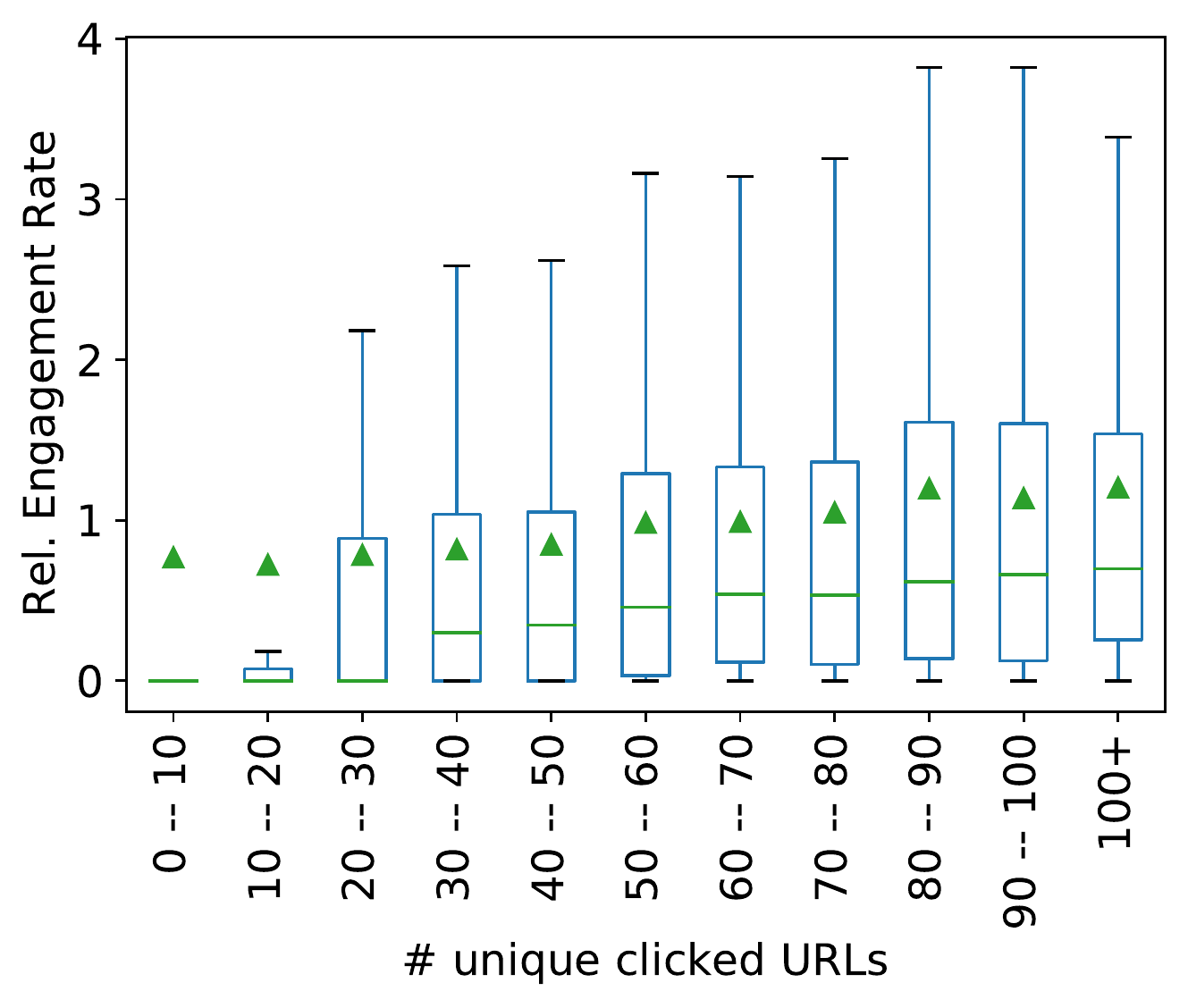}
%   \caption{}
%   \label{fig:box_plot_query_logs:a}
\end{subfigure}%
\begin{subfigure}{.25\textwidth}
  \centering
  \includegraphics[width=\linewidth]{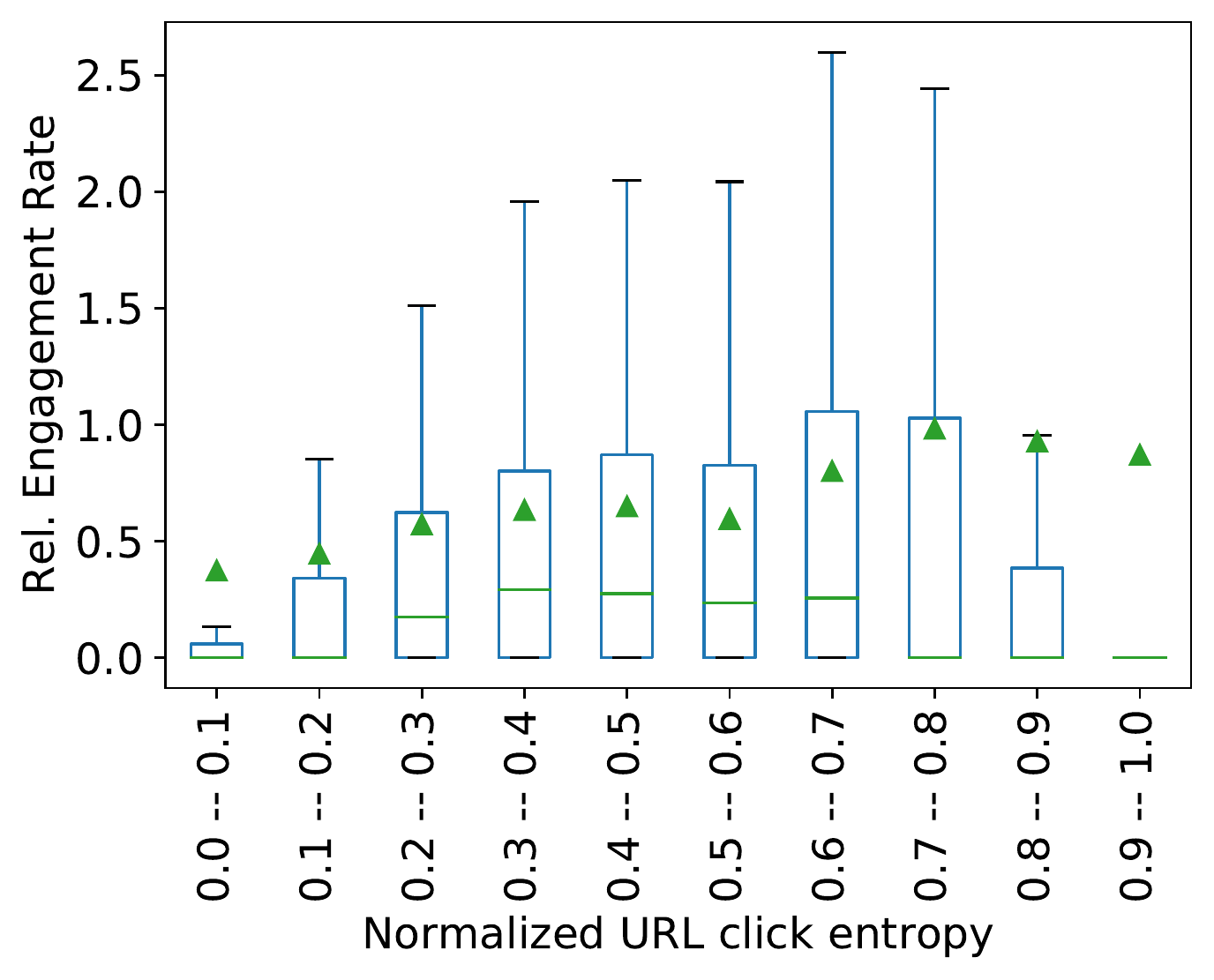}
%   \caption{}
%   \label{fig:box_plot_query_logs:b}
\end{subfigure}%
\vspace{-0.4cm}
\caption{A box plot for the relative engagement rate with respect to (a) the number of unique clicked URLs for the query, and (2) the normalized entropy of click distribution on URLs.}
\label{fig:box_plot_query_logs}
\vspace{-0.6cm}
\end{figure}

% \begin{figure}[t]
%     \centering
%     \includegraphics[width=\linewidth]{img/boxplot_unique_next_queries.pdf}
%     \caption{Box plot of Clarification CTR vs. the number of unique reformulation for the query.}
%     \label{fig:ctr_unique_next_queries}
% \end{figure}

% \begin{figure}[t]
%     \centering
%     \includegraphics[width=\linewidth]{img/boxplot_unique_clicked_urls.pdf}
%     \caption{Box plot of Clarification CTR vs. the number of unique clicked URLs for the query.}
%     \label{fig:ctr_unique_urls}
% \end{figure}

\vspace{-0.2cm}
\subsubsection{Analyzing Clarification Engagement Based on Historical Click Data}
We hypothesize that as the number of aspects for the query increases, the necessity for clarification also increases. To study this hypothesis, we measure the number of aspects per query based on the following criteria:
\begin{itemize}[leftmargin=*]
    % \item Using query reformulation data: for each query $q$ in our data, we looked at a historical query logs and counted the number of unique next queries $q'$ appeared in the same session, such that $q'$ contains $q$. In this case, $q'$ is a more specialized query and we can assume that it demonstrates one aspect of the query. To reduce noise, we only count the $q-q'$ pair, only if it appears in at least two different sessions.
    
    \item Using click data on SERP: for each query $q$ in our data, we looked at a historical click logs and counted the number of unique URLs clicked for the query $q$. %Similar to the query reformulation data, we only consider the clicks with a minimum frequency of 2.
    
    \item Since some clicked URLs may be very related and do not represent different aspects, we follow the approach used in \cite{Koutra:2015} and computed the click distribution entropy normalized by the maximum entropy as an indicator of aspect diversity for the query.
\end{itemize}

The detailed description of click data used in this analysis is presented in Section~\ref{sec:model:iq}.
The results are plotted in \figurename~\ref{fig:box_plot_query_logs} and show that as the number of unique clicked URLs increases the relative engagement rate (both average and median) increases. This is also generally the case when the entropy of click distribution increases. Generally speaking, the unique number of clicked URLs and the click entropy are good indicators of user engagement with clarifying questions.
% that as the number of unique next queries increases up to $60$, the median engagement rate with the \metis also increases. The median for the queries with over $60$ unique next queries are relatively close. The average engagement rate, however, has no correlation with the increase in the number of unique next queries. In the far left bin, the average is even outside the box, which indicates the high number of outliers in this bin. On the other hand, We observe an increase in both average and median engagement rate as the number of unique clicked URLs increases.

\vspace{-0.2cm}
\subsection{Analyzing Clarification Impact and Quality}
\label{sec:analysis:rq4}
In Sections~\ref{sec:analysis:rq2} and \ref{sec:analysis:rq3}, we analyze user interactions with clarification panes in web search. In the next set of analysis, we study the impact of clarification on search experience (i.e., RQ3 in Section~\ref{sec:analysis:rqs}). Since SERP contains multiple elements, such as the result list, the entity card, and the answer box, one cannot simply compute the satisfying click ratio as a full indicator of search satisfaction. \citet{Hassan:2013} shows that measuring user satisfaction can go beyond clicks and for example query reformulation can be used as a signal for user satisfaction. Therefore, because there are multiple SERP elements that can satisfy user satisfaction, we instead focus on dissatisfaction. Clicking on the result list with a small dwell time (i.e., unsatisfying clicks) or reformulating a query with a similar query within a time interval that is short enough (such as five minutes) implies dissatisfaction~\cite{Hassan:2013}. We measured dissatisfaction for the sessions in which users interact with clarification, and observed 16.6\% less dissatisfaction compared to the overall dissatisfaction of the search engine. Note that there are many queries for which the clarification pane is not shown. Therefore, this relative number is \textit{not} a completely representative comparison, however it gives us some idea on the overall impact of clarification on search quality. Since clicking on a candidate answer in clarification leads to a new query and a new SERP, A/B testing for measuring the impact of clarification in search could be also quite challenging here. Some of these challenges have been discussed by~\citet{Machmouchi:2016}. A comprehensive study of user satisfaction while interacting with clarifying questions is left for future work.

We also observe that in 7.30\% of the interactions with the clarification pane, users click on multiple candidate answers. This suggests that in many of these cases, the users would like to \emph{explore} different candidate answers provided by the system. In other words, this observation shows that there is a promise in using clarification with candidate answers for \emph{exploratory search}.

\begin{table}[t]
    \centering
    \vspace{-0.4cm}
    \caption{The human labels for the clarification panes.}
    \vspace{-0.4cm}
    \begin{tabular}{llll}\toprule
        \textbf{Label} & \textbf{\% Good} & \textbf{\% Fair} & \textbf{\% Bad} \\\midrule
        Overall label & 6.4\% & 86.5\% & 7.1\% \\
        Landing page label & 89.1\% & 6.6\% & 4.3\% \\\bottomrule
    \end{tabular}
    \label{tab:human_labels}
    \vspace{-0.5cm}
\end{table}

Another approach to measure the impact of search clarification is measuring search quality using human annotations. To do so, we sampled 2000 unique queries from the search logs and asked three trained annotators to provide labels for each query-clarification pair. Following \cite{Aliannejadi:2019,Zamani:2020:WWW}, we first asked the trained annotators to first skim multiple pages of search results for the query to have a sense on different possible intents of the query. We then asked them to provide the following labels for each clarification pane: 
\begin{itemize}[leftmargin=*]
    \item Overall label: the overall label is given to the whole clarification pane in terms of its usefulness for clarification, comprehensiveness, coverage, understandability, grammar, diversity, and importance order. In summary, they are asked to assign a Good label, if all the mentioned criteria are met. While, the Fair label should be assigned to an acceptable candidate answer set that does not satisfy at least one of the above criteria. Otherwise, the Bad label should be chosen.
    
    \item Landing page quality: the search quality of the secondary SERP obtained by clicking on each candidate answer. A secondary SERP is considered as Good, if the answer to all possible information needs behind the the selected answer can be \emph{easily} found in a prominent location in the page (e.g., an answer box on top of the page or the top three documents) and the retrieved information \emph{correctly} satisfies the possible information needs. If the result page is still useful but finding the answer is not easy, the Fair label should be chosen. Otherwise, the landing page is Bad.
\end{itemize}

% \vspace{-0.2cm}
A detailed description of each label with multiple examples is provided to the annotators. In some rare cases (less than $2\%$), there is no agreement between the annotators (i.e., no label with more than 1 voter). In such cases, we dropped the query-clarification pair from the data. The overall Fleiss' kappa inter-annotator agreement is 72.15\%, which is considered as good. 

The results for human annotations are shown in \tablename~\ref{tab:human_labels}. According to the table, the majority of secondary search results (i.e., landing page) after clicking on each individual option are labeled as Good, so the query intent was addressed in a prominent location of the SERP. For the overall label, most annotators tend to choose Fair as the label. Note that Fair still meets some high standards due to the description provided to the annotators. The reason is that they could mostly argue that there is an intent that is not covered by the \metis, and thus it should not get a Good label.

\begin{figure*}
\vspace{-0.8cm}
\begin{subfigure}{.25\textwidth}
  \centering
  \includegraphics[width=\linewidth]{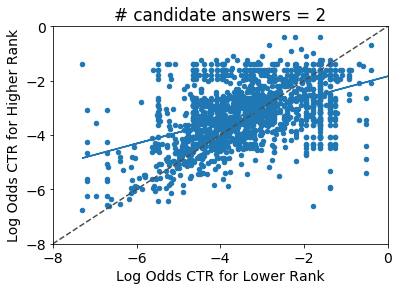}
%   \caption{\# candidate answers = 2}
%   \label{fig:sfig1}
\end{subfigure}%
\begin{subfigure}{.25\textwidth}
  \centering
  \includegraphics[width=\linewidth]{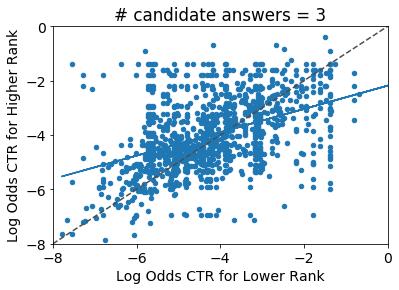}
%   \caption{\# candidate answers = 3}
%   \label{fig:sfig1}
\end{subfigure}%
\begin{subfigure}{.25\textwidth}
  \centering
  \includegraphics[width=\linewidth]{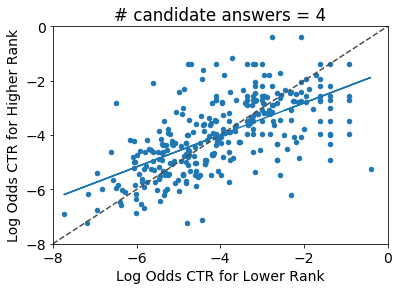}
%   \caption{\# candidate answers = 4}
%   \label{fig:sfig2}
\end{subfigure}%
\begin{subfigure}{.25\textwidth}
  \centering
  \includegraphics[width=\linewidth]{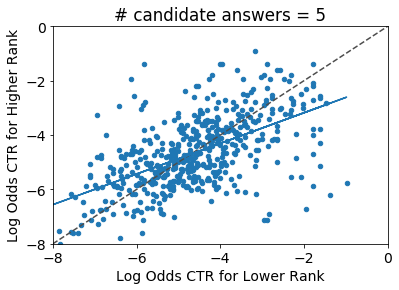}
%   \caption{\# candidate answers = 5}
%   \label{fig:sfig1}
\end{subfigure}%
\vspace{-0.4cm}
\caption{Log odds scatter plot for the click rates of the same candidate answer on the lower position (x axis) and the higher position (y axis) when swapping adjacent candidate answers.}
\label{fig:scatter_position_bias}
\vspace{-0.4cm}
\end{figure*}

\vspace{-0.2cm}
\section{Exploring Click Bias}
\label{sec:bias}
In the last section, we study the engagement rates received by the \metis in web search. In this section, we extend our analysis to the interactions with individual candidate answers. Such analysis would be useful for developing effective models for re-ranking candidate answers or even replacing them. However, implicit feedback could be biased for a number of reasons, such as presentation. \figurename~\ref{fig:ambiguous_vs_faceted} shows that for both query types, the conditional click probability decreases by the increase in the candidate answer position. Note that the candidate answers are presented horizontally in the interface and the first position means the far left candidate answer in \figurename~\ref{fig:bing}. This observation might be due to the fact that the \metis sorts candidate answers based on their popularity and relevance. On the other hand, this could be also due to position and presentation biases in user behaviors. This section provides a preliminary analysis of bias in the click data observed on each candidate answer. 

In the experiments designed for this section, we followed the process used by \citet{Craswell:2008} for studying position bias in web search. In more detail, we created a data set $D$ whose instances are in the form of $(q, C, C')$, where $q$ is a query while $C$ and $C'$ are two difference clarification panes for $q$. We make sure that the clarifying question and the candidate answer set in both $C$ and $C'$ are the same. The only different between $C$ and $C'$ is that \emph{two adjacent candidate answers} are swapped. Therefore, as suggested in \cite{Craswell:2008}, this data allows us to focus on the click distribution on two adjacent candidate answers where their contents and their relevance do not change, while their positions change. This resulted in $46,573$ unique queries and $132,981$ data points in our data.

To study click bias in $D$, we first solely focus on the position. To do so, for each triplet $(q, C, C') \in D$, assume that the candidate answer in position $i$ is swapped with the one in position $i+1$. In other words, $C_i = C'_{i+1}$ and $C_{i+1} = C'_{i}$, where the subscripts show the position of candidate answer (note that $\forall j \neq i, i+1: C_j = C'_j$). We then construct the following two-dimensional data points:
\begin{align*}
    <\text{click rate for~} C_i, ~\text{click rate for~} C'_{i+1}> \\
    <\text{click rate for~} C'_i, ~\text{click rate for~} C_{i+1}>
\end{align*}

These pairs show what would be the click rate on the same candidate answer if it ranks higher for only one position. We repeat this process for all the data points in $D$. The scatter plots for the created data points in a log odds space ($\text{log\_odds}(p) = \log(\frac{p}{1-p})$) are shown in \figurename~\ref{fig:scatter_position_bias}. Note that in a perfect scenario, all points should be on the diagonal in the figures. However, this perfect scenario never happens in practice. We also fit a line (i.e., the solid line) to the data points in each scatter plot to better demonstrate the distribution of data points in this space. As shown in the figure, the slope of the line generally gets closer to the diagonal as the number of options increases. The reason is that as the number of options increases, the click bias in the lower positions are far less than the bias in the higher positions and this influences the overall click bias. 

\begin{table}[t]
    \centering
    \caption{Percentage of points that would receive higher click rate if moved to a higher position (i.e., \% points above the diagonal in \figurename~\ref{fig:scatter_position_bias}). Note that the distance from diagonal is visualized by the line fitted on the data in \figurename~\ref{fig:scatter_position_bias}.}
    \vspace{-0.4cm}
    \begin{tabular}{ccccccccc}\toprule
        \textbf{\# candidate answers} & \textbf{1 $\leftrightarrow$ 2} & \textbf{2 $\leftrightarrow$ 3} & \textbf{3 $\leftrightarrow$ 4} & \textbf{4 $\leftrightarrow$ 5} \\\midrule
        \textbf{2} & 56.34\% \\
        \textbf{3} & 56.17\% & 57.89\% \\
        \textbf{4} & 47.28\% & 57.63\% & 55.62\% \\
        \textbf{5} & 48.50\% & 52.32\% & 53.54\% & 49.77\% \\\bottomrule
    \end{tabular}
    \label{tab:position_bias}
    \vspace{-0.4cm}
\end{table}

\begin{figure*}
% \vspace{-0.8cm}
\begin{subfigure}{.25\textwidth}
  \centering
  \includegraphics[width=\linewidth]{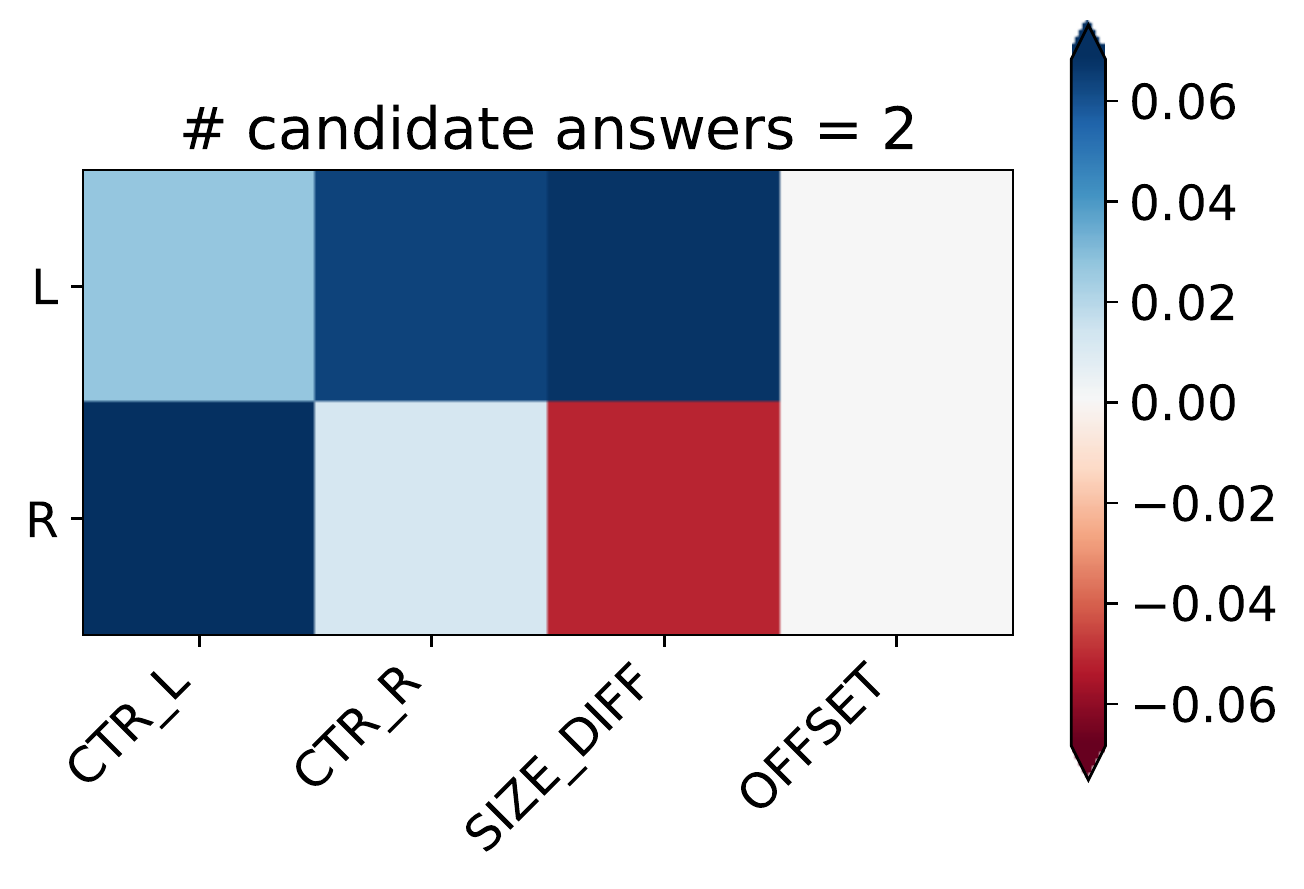}
%   \caption{\# candidate answers = 2}
  \label{fig:sfig1}
\end{subfigure}%
\begin{subfigure}{.25\textwidth}
  \centering
  \includegraphics[width=\linewidth]{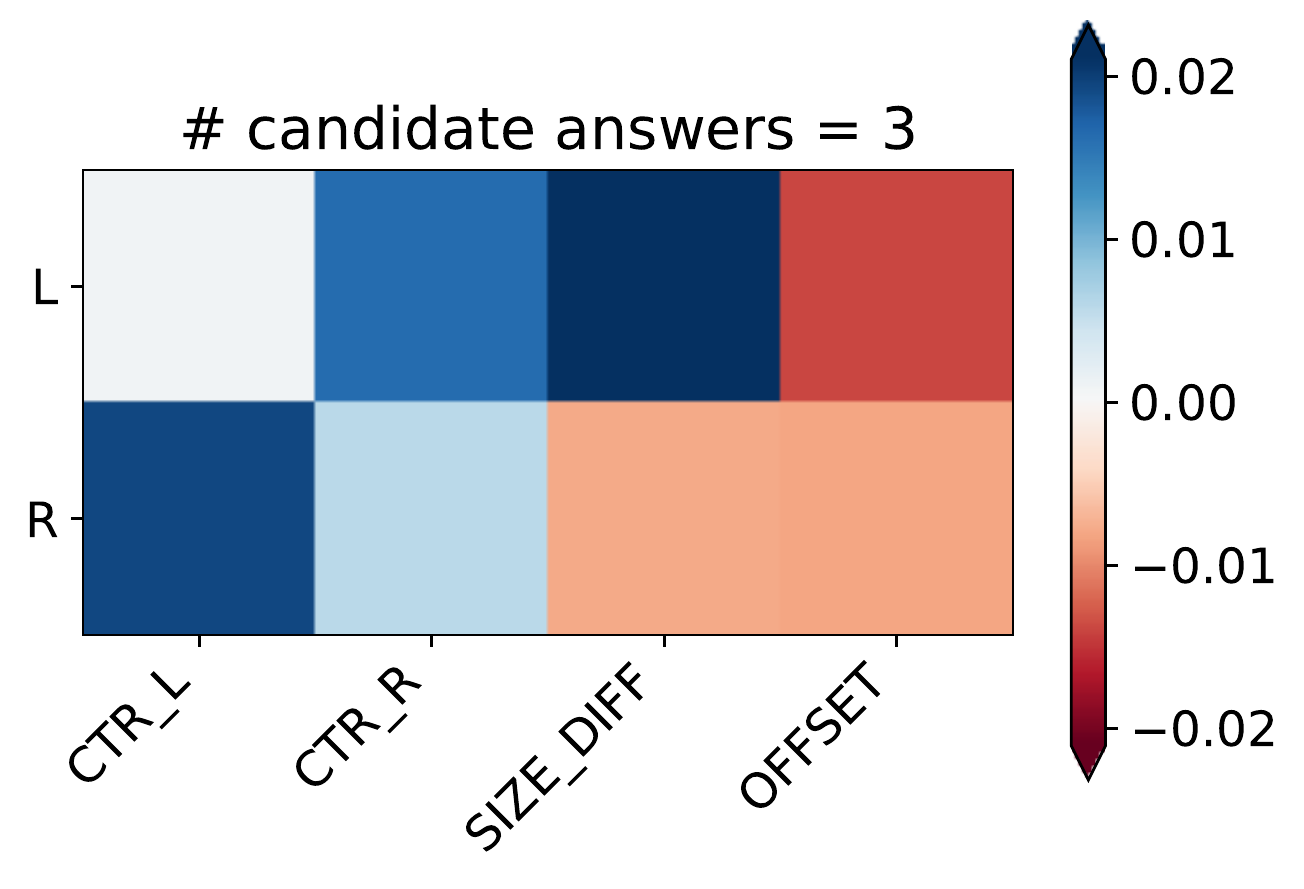}
%   \caption{\# candidate answers = 3}
  \label{fig:sfig1}
\end{subfigure}%
\begin{subfigure}{.25\textwidth}
  \centering
  \includegraphics[width=\linewidth]{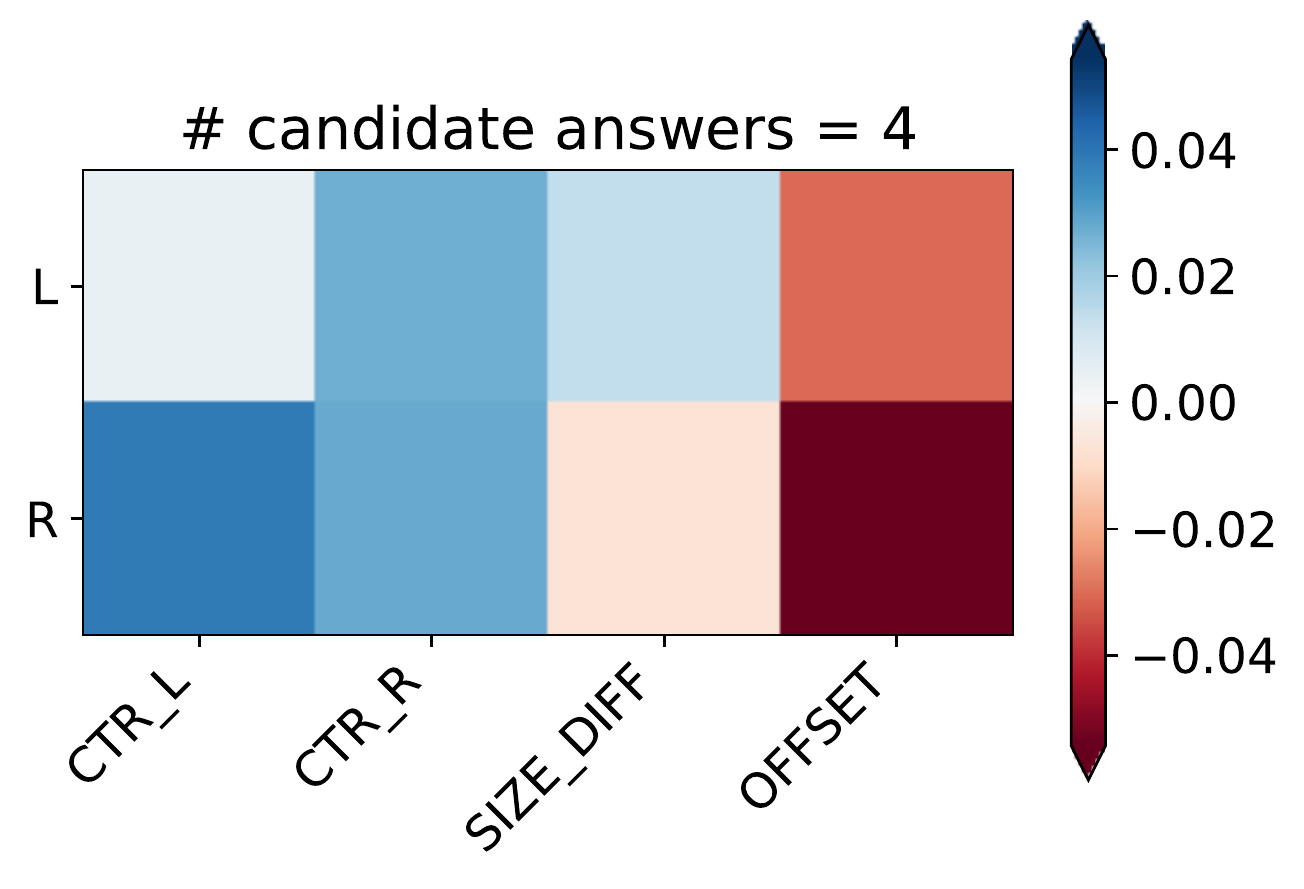}
%   \caption{\# candidate answers = 4}
  \label{fig:sfig2}
\end{subfigure}%
\begin{subfigure}{.25\textwidth}
  \centering
  \includegraphics[width=\linewidth]{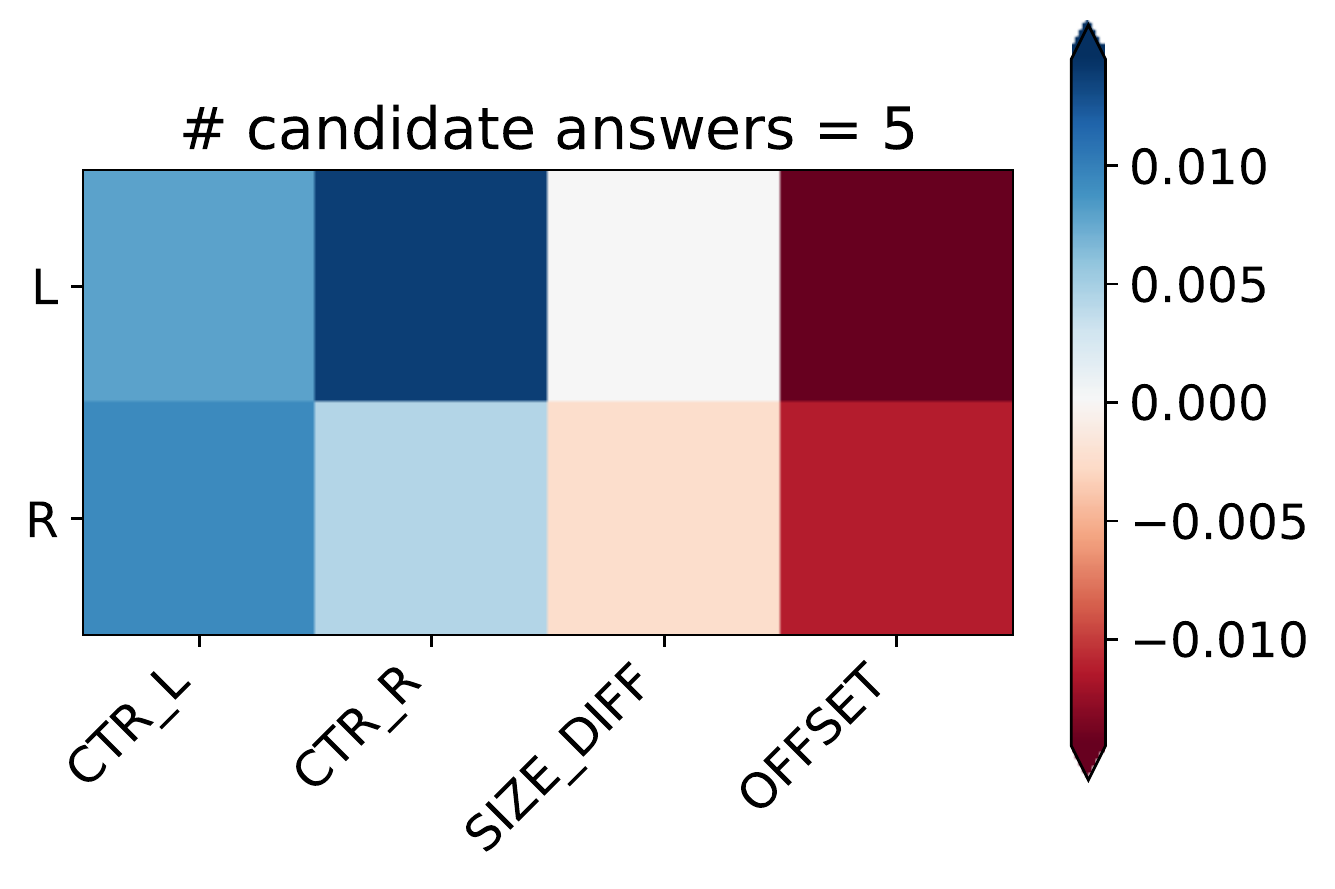}
%   \caption{\# candidate answers = 5}
  \label{fig:sfig1}
\end{subfigure}%
\vspace{-0.8cm}
\caption{Feature weights learned by logistic regression for predicting click rate when two adjacent candidate answers are swapped. The figure should be viewed in color.}
\label{fig:logistic_regression_weights}
\vspace{-0.4cm}
\end{figure*}

\begin{table*}[t]
    \centering
    \caption{Cross entropy for click rate estimation models. Lower cross entropy indicates more accurate click rate estimation.}
    \vspace{-0.4cm}
    \begin{tabular}{lcccc} \toprule
        \textbf{Model} & \textbf{2 options} & \textbf{3 options} & \textbf{4 options} & \textbf{5 options} \\\midrule
        Best Possible & $0.0216 \pm 0.0058$ & $0.0100 \pm 0.0040$ & $0.0097 \pm 0.0049$ & $0.0053 \pm 0.0012$ \\
        Blind click (relevance independent) & $0.1193 \pm 0.0294$ & $0.0604 \pm 0.0275$ & $0.0561 \pm 0.0330$ & $0.0283 \pm 0.0064$ \\
        Baseline (no click bias) & $0.1105 \pm 0.0264$ & $0.0578 \pm 0.0254$ & $0.0539 \pm 0.0329$ & $0.0272 \pm 0.0064$ \\
        Examination & $0.1084 \pm 0.0237$ & $0.0544 \pm 0.0186$ & $0.0517 \pm 0.0260$ & $0.0275 \pm 0.0093$\\
        Cascade & $0.1063 \pm 0.0145$ & $0.0551 \pm 0.0174$ & $0.0510 \pm 0.0189$ & $0.0273 \pm 0.0090$ \\
        Logistic regression & $0.0482 \pm 0.0058$ & $0.0336 \pm 0.0055$ & $0.0333 \pm 0.0064$ & $0.0264 \pm 0.0012$ \\
        \bottomrule
    \end{tabular}
    \label{tab:click_pred}
    \vspace{-0.4cm}
\end{table*}

We also compute the percentage of points above the diagonal in each setting. This shows for what percentage of data points, the same answer with a higher positions would attract more clicks.  The result is reported in \tablename~\ref{tab:position_bias}. Each column in the table shows the position of swapped adjacent answers. The closer the percentage to $50\%$, the less likely there is a click bias. Moreover, all the percentages are typically expected to be higher than or equal to $50\%$, which means options with higher ranks (left) are more likely to be clicked. However, our observation in \tablename~\ref{tab:position_bias} is different. As shown in the table, when the number of answers are $4$ or $5$, the percentage of points above the diagonal is lower than $50\%$ for the $1 \leftrightarrow 2$ setting (this also happens for $4 \leftrightarrow 5$ when the number of candidate answers is $5$, but it is close to $50\%$). The reason for such observation is that position (i.e., rank) is not the only variable that influences click bias. The size of candidate answers also varies as the candidate answer content gets longer (e.g., see \figurename~\ref{fig:bing}). Proper visualization of the click bias considering all of these variables is difficult. Therefore, to study the influence of each variable on click distribution, we train a logistic regression for click prediction. This is similar to the technique used by~\citet{Yue:2010} to study click bias based on different result presentations in web search (e.g., the number of bold terms in the snippets). Therefore, for each triplet $(q, C, C') \in D$, the goal is to predict the click rate for the swapped candidate answers in $C'$ given the observation we had from $C$. We use the following features for the logistic regression model:
\begin{itemize}[leftmargin=*]
    \item CTR\_L: The click rate observed for candidate answer $C_i$.
    \item CTR\_R: The click rate observed for candidate answer $C_{i+1}$.
    \item SIZE\_DIFF: The relative size difference between the candidate answers $C_i$ and $C_{i+1}$. In other words, this feature is equal to $(\text{size}(C_i)-\text{size}(C_{i+1})) / (\text{size}(C_i)+\text{size}(C_{i+1}))$.
    \item OFFSET: The offset of the candidate answer $C_i$. For the first candidate answer, the offset is equal to zero.
\end{itemize}

We train two logistic regressions to predict the following labels:
\begin{itemize}[leftmargin=*]
    \item L: The click rate for candidate answer $C'_i$.
    \item R: The click rate for candidate answer $C'_{i+1}$.
\end{itemize}
Note that the candidate answer $C'_i$ (or $C'_{i+1}$) is in position $i+1$ (or $i$) in $C$. We perform 10 fold cross-validation for training the logistic regression model. The learned feature weights were consistent across folds. The average weights are shown in \figurename~\ref{fig:logistic_regression_weights}. In all the plots, CTR\_L gets a positive weight for the label R and CTR\_R also gets a positive weight for the label L. This shows that the click rate on the same candidate answer in the reverse order is a positive signal for click prediction, which is expected. The weights for two candidate answers shows that the size difference of candidate answers are also very effective in predicting the click bias. As the number of answers increases, the influence of size difference decreases, while the influence of offset increases. The size difference for the label R always gets a negative weight, while this feature gets a positive weight for label L. This is again expected, showing that if we replace the left candidate answer with a larger size answer, the click rate on L would increase, and at the same time the click rate on R would decrease. In other words, the candidate answer size is a strong signal for predicting click rate. The offset has a negative weight for both labels L and R. This suggests that the further the candidate answers from the left, the less likely to observe a click. Note that when the number of candidate answers is two, the offset for all examples is equal to zero and thus it has no effect.

To show that this simple logistic regression predicts the click rate accurately, we compare this model against some simple baselines. The results are reported in \tablename~\ref{tab:click_pred}. Following \citet{Craswell:2008}, we use cross entropy between the true and the predicted click rates as the evaluation metric. The results show that a baseline model that assumes there is no click bias has a much higher cross entropy than the best possible cross entropy (i.e., the entropy of the true labels). The Examination model~\cite{Richardson:2007} and the Cascade model~\cite{Craswell:2008} are user models borrowed from the web search literature. The Examination model assumes each rank has a certain probability of being examined by the user. The Cascade model, on the other hand, assumes that the user views search results from top to bottom, deciding whether to click before moving to the next. Therefore, it also models a skip probability. The assumptions made by both of these models (and many other click models) may not hold in our scenario, where the answers are presented horizontally and their length is small and many of them can be examined by the user at a glance. The results also suggest that these models do not predict the click rate much better than the baseline which assumes there is no click bias. The logistic regression model, however, achieves a much lower cross entropy. Note that the goal of this section is providing some insights into the click bias in the data, and not proposing effective user models for click estimation.

We believe that this preliminary click bias analysis provides some insights into how bias is the user interactions with individual candidate answers. Deeper analyses, for example based on mouse movement and eye-tracking, can shed light on the user click behaviors with clarifying questions and can lead to accurate user models for click estimation and debiasing the data.

\begin{figure*}
\vspace{-0.8cm}
\begin{subfigure}{.33\textwidth}
  \centering
  \includegraphics[width=\linewidth,trim={0 3cm 15.5cm 7.5cm},clip]{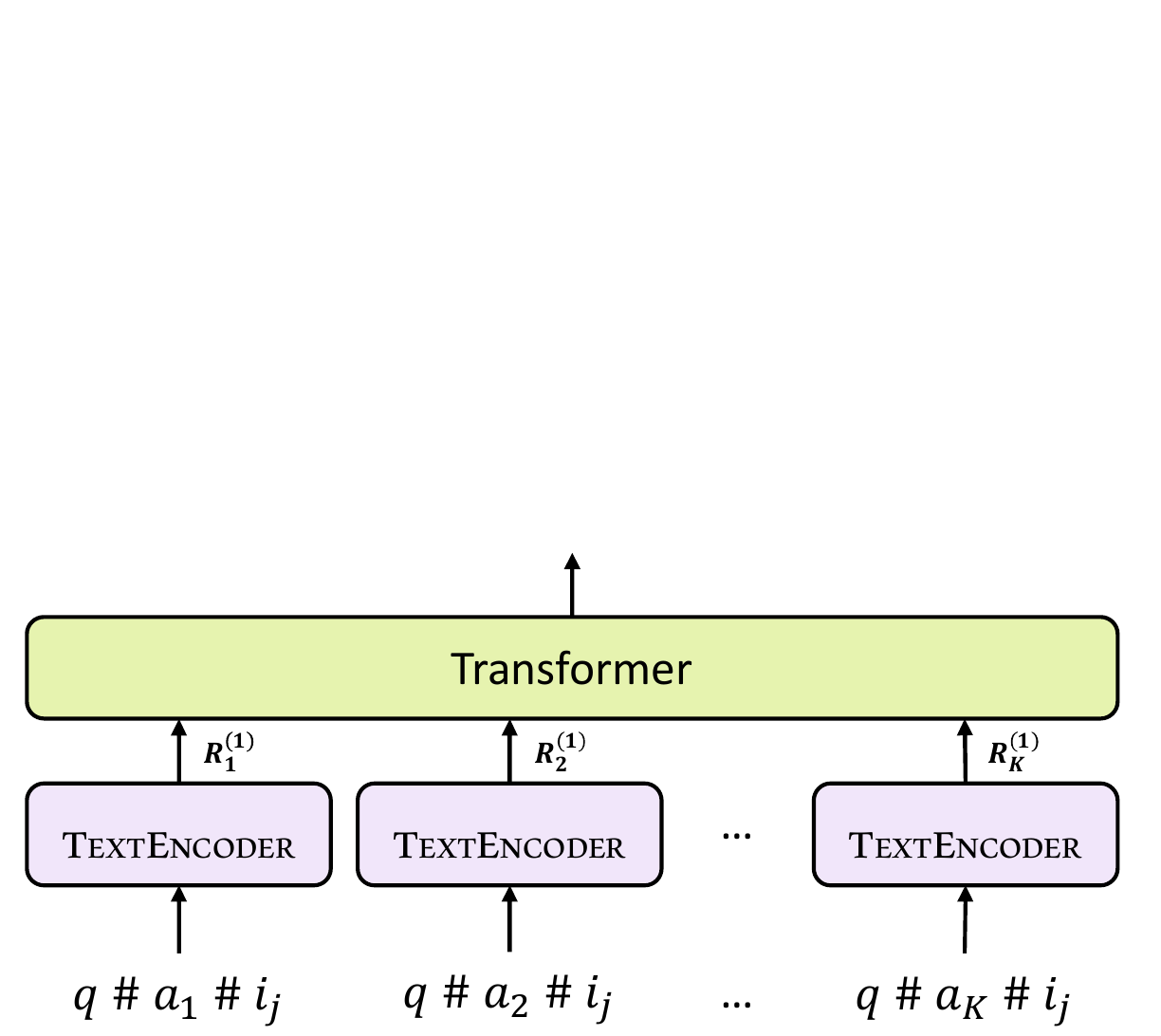}
  \caption{The Individual Intent Encoder model.}
  \label{fig:intent_coverage_encoder}
\end{subfigure}%
\begin{subfigure}{.66\textwidth}
  \centering
  \includegraphics[width=\linewidth,trim={0 3cm 0 5cm},clip]{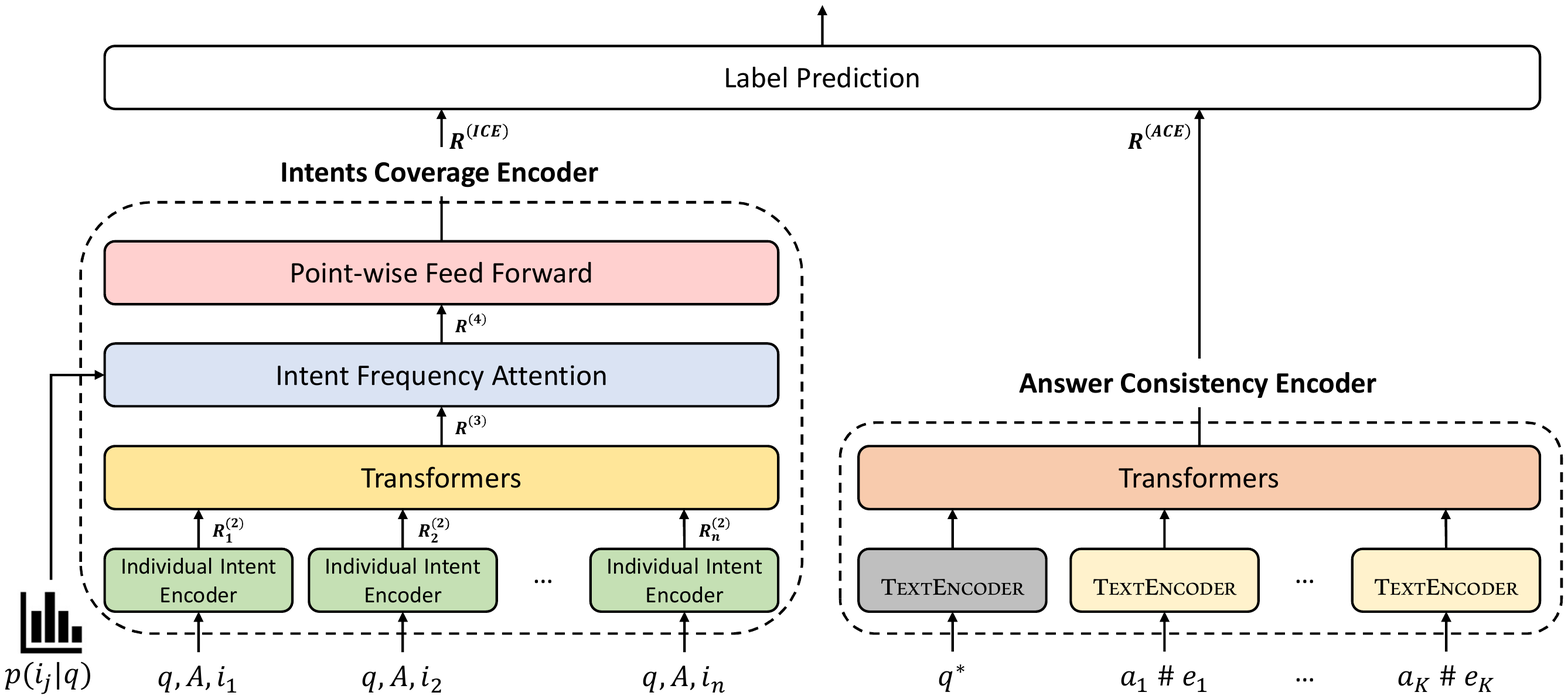}
  \vspace{-0.8cm}
  \caption{The \model architecture.}
%   \label{fig:sfig1}
\end{subfigure}%
\vspace{-.3cm}
\caption{The neural network architecture for \model. Same color indicates shared parameters.}
\label{fig:model}
\vspace{-0.4cm}
\end{figure*}

% \begin{figure*}[t]
%     \centering
%     \includegraphics[width=.7\textwidth,trim={0 4cm 0 5cm},clip]{img/metis-model.pdf}
%     \caption{The neural network architecture for RLC.}
%     \label{fig:model}
% \end{figure*}

% \begin{figure}[t]
%     \centering
%     \includegraphics[width=.8\linewidth,trim={0 3cm 15.5cm 11cm},clip]{img/intent-coverage-encoder.pdf}
%     \caption{Intent Coverage Encoder}
%     \label{fig:intent_coverage_encoder}
% \end{figure}

\vspace{-0.2cm}
\section{Improving Clarification using User Interaction Data}
\label{sec:model}
A fundamental task in search clarification is re-ranking and selecting the clarifying questions generated by different models under different assumptions. A few clarifying question generation models are presented in~\cite{Zamani:2020:WWW}. Based on the analyses presented in Section~\ref{sec:analysis}, we introduce the following features for re-ranking clarification panes in response to a query: (1) question template (a categorical feature), (2) query length, (3) query types (see \tablename~\ref{tab:ctr_per_query_type}), (4) the number of candidate answers, (5) the number of unique clicked URLs, and (6) the URL normalized click entropy. A number of these features are query-specific. To measure how much the clarification pane clarifies different query intents, we can use the Clarification Estimation model presented in \cite{Zamani:2020:WWW}. However, some aspects of clarification (e.g., candidate answer coherency) is missing or is not effectively addressed in this feature. In the following, we propose an end to end neural model to fill these gaps. The model is mainly trained based on user interaction data and further fine-tuned using a small set of human labeled data.

% In this section, we introduce \model, a neural model for learning a representation for a \metis, i.e., a clarifying question together with its candidate answers. We design our model solely based on attention, due to its efficiency and effectiveness in various NLP and IR tasks~\cite{Devlin:2019,Vaswani:2017,Nogueira:2019}. \model is mainly trained based on the user click data as implicit feedback. The learned representations can be used for selecting (or re-ranking) clarifying questions for a given query.

Let us first introduce our notation. Let $T$ denote a training set containing triplets of $(q, C, L)$, where $q$ is a unique query, $C=[c_1, c_2, \cdots, c_m]$ is a set of $m$ clarification panes for the query, and $L = [l_1, l_2, \cdots, l_m]$ is the labels associated with the clarification panes. Each \metis $c_j$ includes a clarifying question $q^*$ and a list of $K$ candidate answers $A=[a_1, a_2, \cdots, a_K]$, where $K=5$ in our setting. Additionally, let $I_q$ denote the intent set for the query $q$ with $n$ intents, whose $j$\textsuperscript{th} element is a pair $(i_j, w_j)$, where denotes an intent ($i_j$) and its weight ($w_j$). Note that the query intent set is often unknown to a system, but there exist few approaches for estimating the intent set based on query logs and click data. We later explain how we built the intent set $I_q$ for our experiments (See Section~\ref{sec:model:iq}). The goal is to train a representation learning model for each query-clarification pair. This model can be used for selecting or re-ranking clarification panes. 

\vspace{-0.2cm}
\subsection{Representation Learning for Clarification}
\label{sec:model:network}
We design our neural model based on the following assumptions:

\noindent \textbf{Assumption 1.} A good \metis should clarify different intents of the query, in particular the most frequent intents.

\noindent \textbf{Assumption 2.} The candidate answers in a good \metis should be coherent and also consistent with the clarifying question.

Assumption 1 is indirectly related to the analysis done in \figurename~\ref{fig:box_plot_query_logs}, which shows that queries with more unique clicked URLs would lead to higher engagement rates. This shows that covering a wide range of intents is an important factor in clarifying questions, which leads us to the first assumption. Given these assumptions, our model, called \model,\footnote{stands for Representation Learning for Clarification.} is built based on two major components: \textit{Intents Coverage Encoder} and \textit{Answers Consistency Encoder}. The architecture of \model is depicted in \figurename~\ref{fig:model}.

\begin{table}[t]
    \centering
    % \vspace{-0.2cm}
    \caption{The training and test data used in our experiments.}
    \vspace{-0.4cm}
    \begin{tabular}{lccc}\toprule
        \textbf{Data} & \textbf{\specialcell{\# training\\queries}} & \textbf{\specialcell{\# test\\ queries}} & \textbf{\specialcell{\# clarifications\\per query}} \\\midrule
        Click data & 137,392 & 3925 & 6.2\\
        Labeled data & 1848 & 122 & 10\\\bottomrule
    \end{tabular}
    \label{tab:exp_data}
    \vspace{-0.5cm}
\end{table}

\vspace{-0.2cm}
\subsubsection{Intents Coverage Encoder}
This component learns a high-dimensional vector representing the intent coverage of the candidate answer set. We first create $K \times n$ triplets $(q, a_k, i_j)$ for $1 \leq k \leq K$ and $1 \leq j \leq n$. For each of these triplets, we create a sequence \texttt{<b> query <s> answer <s> intent <e>} with some boundary tokens and feed the sequence to a text encoder network for obtaining the representation $R^{(1)}_{kj} = \text{\sc{TextEncoder}}(q, a_k, i_j)$. See Section~\ref{sec:model:bert} for more information on {\sc{TextEncoder}}. Next, we would like to see whether each intent is covered by the candidate answer set. Therefore, we concatenate all the representations $R^{(1)}_{kj}$ for all $1 \leq k \leq K$ and feed the obtained vector to a Transformer encoder, which consists of multiple Transformer layers~\cite{Vaswani:2017}. The self-attention mechanism in Transformer helps the model learn a representation for the coverage of the $j$\textsuperscript{th} intent by the answer set. This results in $n$ representations $R^{(2)}_{j}$, one per query intent. 

Different query intents may be related, especially since they are automatically estimated using some algorithms. Therefore, we apply a Transformer Encoder layer on top of all individual intent representations, whose self-attention mechanism would lead to learning accurate representations for related intents. This layer gives us $R^{(3)}_{j}$ for each intent $i_j$. In addition, some intents are more common than the others. According to Assumption 1, we expect the model to particularly cover those common intents. Therefore, we use the intent weights as attentions for intent coverage representation. Formally, $R^{(4)}_{j} = \frac{w_j}{\sum_{j'}{w_{j'}}} R^{(3)}_{j}$. This layer is followed by two point-wise feed-forward layers to adjust the representation space and add non-linearity. This component returns the intent coverage encoding $R^{(ICE)}$.

\begin{table*}[t]
    \centering
    \vspace{-0.6cm}
    \caption{Experimental results for re-ranking clarification panes for a query. The superscripts 1/2/3 indicate statistically significant improvements compared to Clarification Estimation/BERT/LambdaMART without \model, respectively.}
    \vspace{-0.4cm}
    \begin{tabular}{ll:lll:ccc}\toprule
        \multirow{2}{*}{\textbf{Method}} & \multicolumn{1}{c:}{\textbf{Click Data}} & \multicolumn{3}{c:}{\textbf{Labeled Data}} & \multicolumn{3}{c}{\textbf{Landing Pages Quality}} \\
         & \textbf{Eng. Rate Impr.} & \textbf{nDCG@1} & \textbf{nDCG@3} & \textbf{nDCG@5} & \textbf{\%Bad} & \textbf{\%Fair} & \textbf{\%Good} \\\midrule
        Clarification Estimation~\cite{Zamani:2020:WWW} & -- & 0.8173 & 0.9356 & 0.9348 & 11.68\% & 13.24\% & 75.08\%\\
        BERT~\cite{Devlin:2019} & 25.96\%$^{1}$ & 0.8515$^{1}$ & 0.9449 & 0.9425 & 10.52\% & 17.24\% & 72.24\%\\
        LambdaMART w/o \model & 67.27\%$^{12}$ & 0.9001$^{12}$ & 0.9584$^{1}$ & 0.9565$^{1}$ & 5.21\% & 19.45\% & 75.34\% \\\hdashline
        \model & 92.41\%$^{123}$ & 0.9312$^{123}$ & 0.9721$^{123}$ & 0.9702$^{123}$ & 5.63\% & 12.33\% & 82.04\% \\ 
        LambdaMART w/ \model & \textbf{106.18\%$^{123}$} & \textbf{0.9410$^{123}$} & \textbf{0.9822$^{123}$} & \textbf{0.9767$^{123}$} & 4.94\% & 10.21\% & 84.85\% \\
         \bottomrule
    \end{tabular}
    \label{tab:results}
    \vspace{-0.4cm}
\end{table*}

\vspace{-0.2cm}
\subsubsection{Answers Consistency Encoder}
This component focuses on the clarifying question and its answer set. Answer entity types are found useful for generating clarifying questions~\cite{Zamani:2020:WWW}. Therefore, in this component, we first learn a representation for each candidate answer $a_k$ based on the answer text and its entity type (denoted as $e_k$) if exists, concatenated using a separation token and fed into {\sc{TextEncoder}}. We also feed the clarifying question to the {\sc{TextEncoder}}. This results in $K+1$ representations. We further apply a Transformer encoder whose self-attention mechanism helps the model identify coherent and consistent answers. In other words, the attention weights from each candidate answer to the others as well as the question help the model observe the similarity of answers and their entity types. The use of entity type would increase generalization and entity similarity better represents the answer coherency. $R^{(ACE)}$ is the output of this component.

\vspace{-0.2cm}
\subsubsection{Label Prediction}
For the label prediction sub-network, we simply concatenate $R^{(ICE)}$ and $R^{(ACE)}$ and feed the obtained vector to a feed-forward network with two layers. The output dimensionality of this component is 1, which indicates the final score for the given query-clarification pair. 

\vspace{-0.2cm}
\subsubsection{\sc{TextEncoder}}
\label{sec:model:bert}
As mentioned above, each major component in the network starts with a {\sc{TextEncoder}}. There are several approaches for implementing this component. In this paper, we use BERT~\cite{Devlin:2019} -- a Transformer-based network pre-trained on a masked language modeling task. BERT has recently led to significant improvements in several NLP and IR tasks~\cite{Devlin:2019,Nogueira:2019,Padigela:2019}. We use BERT-base which consists of 12 layers, 768 representation dimensions, 12 attention heads, and 110M parameters.\footnote{The pre-trained models can be found at \url{https://github.com/google-research/bert}.} The BERT parameters are fine-tuned in our end-to-end training. The components with the same color in \figurename~\ref{fig:model} share parameters. Note that the {\sc{TextEncoder}} functions with different colors still share the embedding layer (i.e., the first layer), while their attention weight matrices are different and learned for the specific input type.

\vspace{-0.2cm}
\subsubsection{The Intent Set $I_q$}
\label{sec:model:iq}
% \paragraph{\textbf{Query Reformulation Data and Click Data:}} 
We use two datasets for estimating the intents of each query.\footnote{Therefore, there are two Intents Coverage Encoders whose outputs are concatenated.} The first one is the query reformulation data and the second one is click data on documents. These two datasets were obtained from the Bing query logs, randomly sub-sampled from the data collected in a 2 year period of the EN-US market. The query reformulation data is a set of triplets $(q, q', w)$, where $w$ is the frequency of the $q \xlongrightarrow{} q'$ query reformulation in the same session. We use the reformulations in which $q'$ contains $q$ as an estimation for query intent. A similar assumption has been made in~\cite{Zamani:2020:WWW}. From the click data, we use the title of the clicked URLs as an additional source for estimating query intents. We only kept the query reformulations and clicks with a minimum frequency of 2. 

% We try the following models:
% \begin{itemize}[leftmargin=*]
%     \item BERT~\cite{Devlin:2019}: BERT is a Transformer-based language model pre-trained for the language modeling task. BERT has been recently employed in several state-of-the-art NLP and IR tasks~\cite{Devlin:2019,Nogueira:2019,Padigela:2019}. We use the BERT-base which consists of 12 Transformer layers.
    
%     \item DistilBERT~\cite{Sanh:2019}: Despite its effectiveness, BERT suffers from relatively low efficiency and high memory cost. DistilBERT is a smaller version of BERT, trained using knowledge distillation, which performs much faster than BERT without losing substantial effectiveness.
    
%     \item Training Transformer encoders from scratch: This {\sc{TextEncoder}} model is a two layer Transformer network~\cite{Vaswani:2017} with random parameter initialization, with no pre-training. This text encoder is the more efficient than the others.
% \end{itemize}

% Note that for all the above models, the parameters are trained or fine-tuned in an end-to-end training. Note that the first layer in these {\sc{TextEncoder}} models (i.e., the embedding layer) is shared among all parts of the network. The rest of the parameters are sub-network specific. In other words, the clarifying question model uses different {\sc{TextEncoder}} attention weights than the candidate answer representation, however, all candidate answers share the same text encoding parameters.

\vspace{-0.2cm}
\subsection{Training}
\label{sec:model:training}
We train our model using a pair-wise loss function. For two clarification panes for the same query, we get the score from \model and use the softmax operator to convert the scores to probabilities. We use the binary cross entropy loss function for training, i.e., the label for the clarification pane with higher engagement rate is 1. %We train our model using click data and use engagement rate as label for each clarification pane. 
We further fine tune the model using a small set of human labeled data. We optimize the network parameters using Adam with $L_2$ weight decay, learning rate warm-up for the first $5000$ steps and linear decay of the learning rate. The learning rate was set to $10^5$. In the following, we introduce our datasets:

% \begin{table*}[t]
%     \centering
%     \caption{Experimental results.}
%     \begin{tabular}{lccc}\toprule
%         \multirow{2}{*}{\textbf{Method}} & \textbf{Click data} & \multicolumn{2}{c}{\textbf{Labeled data}} \\
%          & \textbf{Engagement rate} & \textbf{Overall label} & \textbf{Average answer label} \\\midrule
%         Lower bound & 0.0008 & 1.9016 & 1.0872 \\
%         Upper bound & 0.3845 & 2.3196 & 1.6702 \\\midrule
%         Random & 0.0903 & 2.1122 & 1.3821 \\
%          \bottomrule
%     \end{tabular}
%     \label{tab:results}
% \end{table*}

% \section{Experiments}
% \label{sec:exp}
% In this section, we evaluate the proposed neural model for the \metis selection task.
% \vspace{-0.2cm}
% \subsection{Training and Evaluation Data}
% \label{sec:model:data}
\vspace{-0.2cm}
\paragraph{\textbf{Clarification Click Data:}}
From the data described earlier in \tablename~\ref{tab:metis_data}, we kept clarifying questions with at least 10 impressions, and at least two different clarification panes that have different engagement rates, i.e., click rates. We split the data randomly into train and test based on the queries. For more details, see \tablename~\ref{tab:exp_data}. %The average number of candidate clarification panes per query is 6.2 and 10 in the click data and labeled data, respectively.

\vspace{-0.2cm}
\paragraph{\textbf{Clarification Labeled Data:}}
We obtained an overall label for clarification and the secondary search result page (landing page) quality labels using the instructions mentioned in Section~\ref{sec:analysis:rq4}. We split the data into train and test sets and no query is shared between the sets. The statistics of this data is also reported in \tablename~\ref{tab:exp_data}. Note that in the labeled data we re-rank 10 clarifying questions per query. If the number of labeled clarifying questions are less than 10, we randomly add negative samples with label $0$ from the clarifying questions for other queries.

\vspace{-0.2cm}
\paragraph{\textbf{Entity Type Data:}} For answer entity types, we used an open information extraction toolkit, i.e., Reverb~\cite{Fader:2011}, to extract ``is a'' relations from a large-scale corpus (over 35 petabyte of search snippets). We only kept the relations with the confidence of at least $96\%$. This results in over 27 millions relations for over 20 millions unique phrases. The data contains over 6 millions entity types. %, out of which over 17,000 entity types have a minimum frequency of 10. This data covers the entity types for over $40\%$ of search queries.

% \subsection{Experimental Setup}
% We use ADAM (Kingma & Ba, 2014) with the initial learning rate set to 3 × 10−6
% , β1 = 0.9,
% β2 = 0.999, L2 weight decay of 0.01, learning rate warmup over the first 10,000 steps, and linear
% decay of the learning rate. We use a dropout probability of 0.1 on all layers.

% \hamed{parameter setting}

% \subsubsection{Evaluation Metrics}

\vspace{-0.3cm}
\subsection{Clarification Re-Ranking Results}
We first trained the model using $90\%$ of the training set and use the remaining $10\%$ for hyper-parameter tuning of all models, including the baselines. Once the hyper-paraters were selected, we trained the final model on the whole training set and computed the result on the test set.
The results for the proposed method and some baselines are reported in \tablename~\ref{tab:results}. For the click data, we re-rank the clarification panes and select the first one and report the engagement rate. We finally compute the average engagement rates across queries. The engagement rates are reported relative to the performance of Clarification Estimation~\cite{Zamani:2020:WWW}. The BERT model uses all the inputs we used in \model, i.e., the query, the clarification pane and the estimated intents. All of these inputs are concatenated using separation tokens and fed to BERT-base with different segment embeddings. %We use the BERT-base model.
LambdaMART~\cite{Burges:2010} w/o \model uses all the features described earlier in Section~\ref{sec:model} plus the BERT-base output. The results show that the proposed method outperforms all the baselines. According to the paired t-test with Bonferroni correction, the improvements are statistically significant ($p\_value < 0.05$). The best model (i.e., LambdaMART w/ \model) achieves an nDCG@1 of $0.9410$. 

\vspace{-0.3cm}
\section{Conclusion and Future Directions}
\label{sec:conclusion}
In this paper, we provided a thorough analysis of large-scale user interactions with clarifying questions in a major web search engine. We studied the impact of clarification properties on user engagement. We further investigated the queries for which users are more likely to interact with the clarification pane. We also explored the impact of clarification on web search experience, and analyzed presentation bias in user interactions with the clarification panes in web search. Our preliminary analysis on click bias showed that users are often intended to click on candidate answers in higher positions and with larger size. Motivated by our analysis, we proposed a set of features and an end to end neural model for re-ranking clarifying questions for a query. The proposed models outperform the baselines on both click data and human labeled data. 

In the future, we intend to study click models for clarification panes to reduce the impact of click bias in ranking candidate answers. We would like to explore user interactions with clarification in devices with limited bandwidth interfaces, such as mobile phones with a focus on speech interactions. Multi-turn clarification is also left for future work.

\vspace{-0.4cm}

% \bibliographystyle{ACM-Reference-Format}
% \bibliography{sigproc} 


% 
% 

\end{document}